\documentclass[aps,prl,twocolumn,floatfix]{revtex4-1}
\usepackage{latexsym}
\usepackage{graphicx}

\newcommand{\figw}{2.2 in}

\usepackage{subfigure}    
\usepackage{epsfig}
\usepackage{float}
\usepackage{hyperref}

\usepackage{amsmath}

\begin{document}

\title{Importance of electron correlations in understanding the
  photo-electron spectroscopy and the Weyl character of
   MoTe$_2$}
\author{Niraj Aryal$^{(1)}$}
\author{ Efstratios Manousakis$^{(1,2)}$}
\affiliation{
$^{(1)}$ Department  of  Physics and National High Magnetic Field Laboratory,
  Florida  State  University,  Tallahassee,  FL  32306-4350,  USA\\
$^{(2)}$Department   of    Physics,   University    of   Athens,
  Panepistimioupolis, Zografos, 157 84 Athens, Greece
}
\date{\today}

\begin{abstract}
  We study the  role of
  electron  correlations  in the  presumed  type  II Weyl  semimetallic
  candidate $\gamma$-MoTe$_2$ by employing density functional
  theory (DFT)  where the on-site Coulomb repulsion
  (Hubbard U) for the Mo 4$d$ states is included within the DFT+U scheme.
  We show that  pure  DFT  calculations  fail  to  describe  important
  features  of  the  light-polarization   dependence  of  the  angular
  resolved  photoemission  intensity which  can  be  accounted for  by
  including the  role of the Hubbard U.  At  the same
  time while pure DFT calculations  cannot  explain
  the angular dependence  of the Fermi surface as  revealed by quantum
  oscillation  experiments (a fact which had raised doubt about the presence
  of the Weyl physics in $\gamma$-MoTe$_2$)
  inclusion  of  such  on-site  Coulomb
  repulsion can.   We find  that while the number of Weyl points (WPs)
  and their  position in the Brillouin  Zone change as a function of U,
  a pair of such WPs very close to the Fermi level survive the
  inclusion of these important corrections. Our calculations
  suggest that
  the Fermi surface of $\gamma$-MoTe$_2$ is in the vicinity of a
  correlations-induced Lifshitz
  transition  which  can be probed experimentally and its interplay with the
  Weyl physics might be intriguing.
\end{abstract}

\maketitle
Transition metal dichalcogenides (TMDs) have 
continued to surprise physicists and chemists for decades by providing an avalanche of 
materials with intriguing chemical, mechanical, electronic and optical properties, 
of both fundamental and technological implications\cite{TMDsWangnature2012,TMDsWilsonAiP1969,TMDsMattheissPRB1973}.
Among several other properties of fundamental importance in physics,
more recently, 
DFT calculations\cite{Wang_Bernevig_2016,TypeIISoluyanov2015,Prediction_ChangNature_2016}
predict that the lesser known inversion-symmetry-breaking TMDs
$\gamma-$WTe$_2$ and $\gamma-$MoTe$_2$ host
Lorentz invariance violating type-II Weyl Fermions.
In a similar fashion to their type-I counterparts (namely, (Ta,Nb)(As,P)), these materials are predicted to host Weyl nodes (WN) at the boundaries
of the electron and hole pockets as well as topological Fermi arcs which connect
Weyl nodes of opposite chiralities\cite{FermiarcsTaAs_XuScience2015,FermiarcsTaAs_HuangNature2015,FermiarcsNbAs_XuNature2015,Armitage_Vishwanath_RMP2018}.
Different anomalies in the transport experiments, such as 
extremely high carrier mobility\cite{UltraHighMobilityNbPWangPRB2016,XLMRNbpShekhar2015}, and chiral anomaly induced negative longitudinal magnetoresistance\cite{ChiralAnomalyTaAs_HuangPRX2015}  are considered 
indirect evidence of Weyl Fermions.

Angular resolved photoemission spectroscopy (ARPES)  is undoubtedly a direct and widely accepted probe of the above mentioned
features of the electronic structure and it has successfully identified 
type-I Weyl candidates by directly imaging the WNs and Fermi arc states\cite{FermiarcsTaAs_XuScience2015,FermiarcsTaAs_HuangNature2015}.
However, in materials which are candidates for realizing type-II Weyl Fermions, ARPES experiments are not as convincing and unambiguous
as in the type-I case
because of the coexistence of the bulk 
electron and hole pockets with surface arc-states and the presence of both trivial and non-trivial Fermi
arcs\cite{Bruno_Baumberger_2016,MinimalModelWeyl_Trivedi__RB2017}.
Nevertheless, in the candidate material $\gamma$-MoTe$_2$ several ARPES studies have claimed to observe Weyl points (WPs) and non-trivial Fermi arcs  in agreement with the DFT calculations
~\cite{MoTe2Arpes_Kaminski_Nature2016,MoTe2Arpes_Zhou_Nature2016,MoTe2Arpes_jiang_Nature2017,Tamai_Baumberger_2016}.
On the other hand, the predicted electron and hole pockets from DFT calculations, which give rise to the WPs and non-trivial topology, 
fail to explain the experimentally observed quantum oscillation (QO) frequencies \cite{MoTe2Aryal2017} in MoTe$_2$. This failure of the DFT calculations
to explain the QO frequencies has raised
a certain degree of doubt about the existence of these WPs, given the
fact that they were the predictions of such calculations.
Also, the high sensitivity of the WPs (of their number, their location and even of their presence)
to the slight change in the lattice parameters \cite{Wang_Bernevig_2016,MoTe2YanPRB2015},
the fact that even the presence of Fermi arcs does not necessarily imply the existence of the WPs and non-trivial topology \cite{Bruno_Baumberger_2016,TrivialMoTe2_Crepaldi_Grioni_PRB2017} 
and the approximate way of incorporating the exchange-correlation effects in the DFT, 
upon which all the predictions about Weyl materials are based, has made this field both exciting and controversial.
A significant amount of effort and resources have been devoted
by the scientific community to understand and possibly use these Weyl and other topological materials in novel electronics, therefore,
it is very important to make sure that their characterization, especially
with regard to these fundamental properties is accurate.

In Ref.~\onlinecite{MoTe2Aryal2017}, the authors have empirically shown
that in order to make any sensible
comparison between the experimentally observed QO
frequencies
and the DFT calculated ones for the Weyl candidate MoTe$_2$, 
the relative band energies of the electron and hole pockets have to be shifted in opposite directions by a constant momentum-independent amount of energy ($\sim 50$ meV). 
These authors also showed that such a shift was enough to explain 
the ARPES observed band structure.
This adjustment of the band structure, however, sheds doubt about the existence
of the Weyl points whose presence depends strongly on small amounts of
energy displacement of the hole and electron bands.

In this letter, we resolve the above described discrepancy
by studying the electronic structure of MoTe$_2$ by means of the DFT+U method.
Our results show that the ``ad hoc'' shifting of the bands attempted in Ref.~\onlinecite{MoTe2Aryal2017} is not only empirical but also justified.
The electron and hole bands, predicted to form WPs, are
the result of hybridization of Mo-$d$ and Te-$p$ orbitals (see Fig.~1 of the Supplemental Material(SM)~\onlinecite{Supplementary}).
Hence, as expected introduction of a Hubbard U term on the Mo $d$-orbital to take into account the effects of Coulomb correlations arising from the more
localized nature of this orbital would lead to different contributions to the energy of these bands.
A reasonable value of Hubbard U ($\approx 3 eV$) provides  a  better agreement
with the QO and ARPES data. In addition, we demonstrate that in order to
explain the polarization dependence of the ARPES results we need
to introduce  a non-zero U.
More importantly, while many other pairs of Weyl points vanish by increasing the
value of U an interesting Weyl pair very near the Fermi level is found
to survive.
Also, our calculations indicate that the system is in close vicinity to a Lifshitz transition and the WPs are much closer to the Fermi level
than predicted by DFT calculations and these predictions can be explored
experimentally.

\begin{figure}[htb]
    \begin{center}
\vskip 0.08 in
\hskip -0.02 in
        \subfigure[]{
            \includegraphics[width=1.6 in]{Fig1a.eps}
            \label{fig:U_3_bands}
	}
\hskip 0.01 in
        \subfigure[]{
            \includegraphics[width=1.30 in]{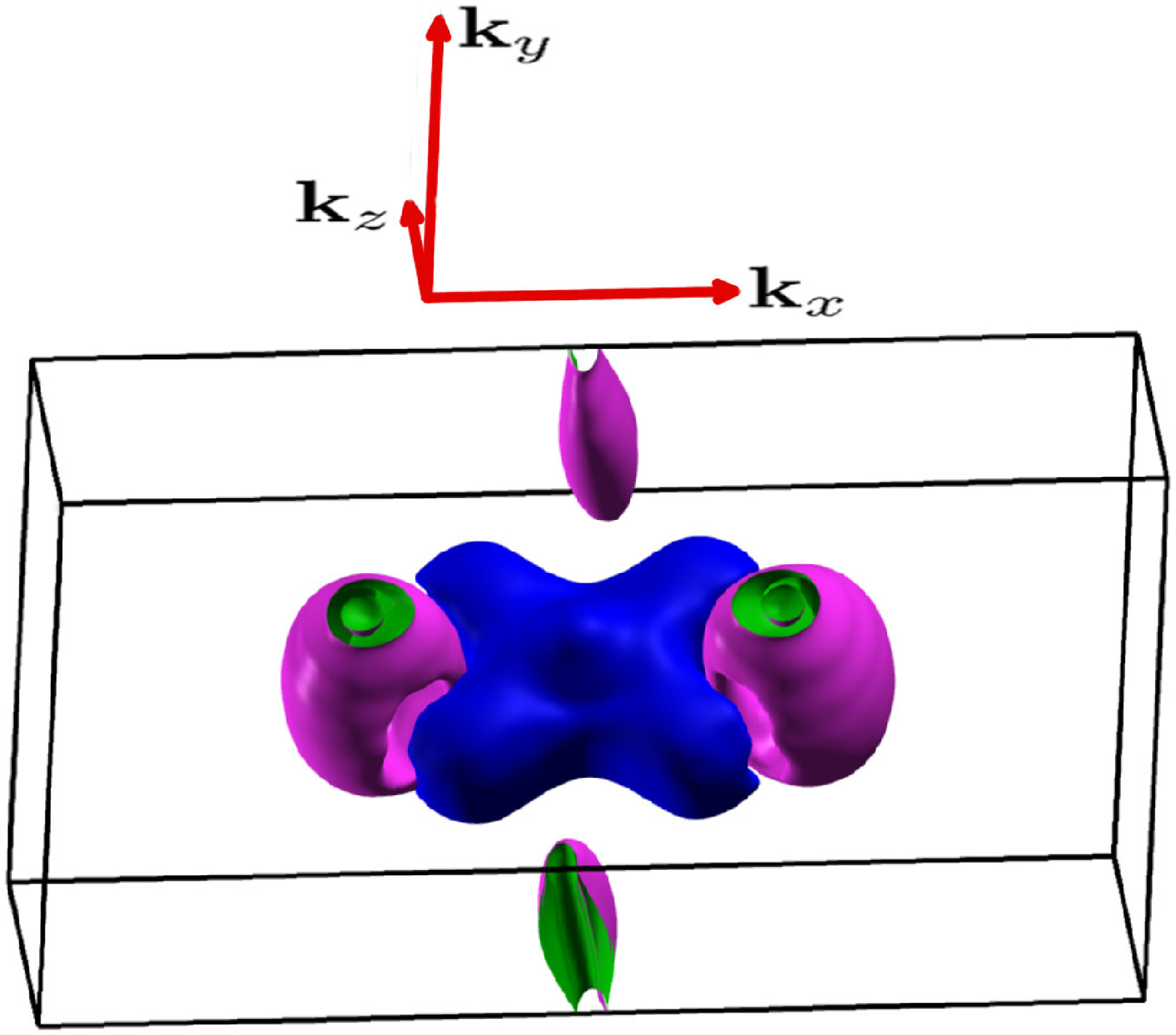}
            \label{fig:FS_unshifted}
	}
\vskip 0.01 in
        \subfigure[]{
            \includegraphics[width=1.33 in]{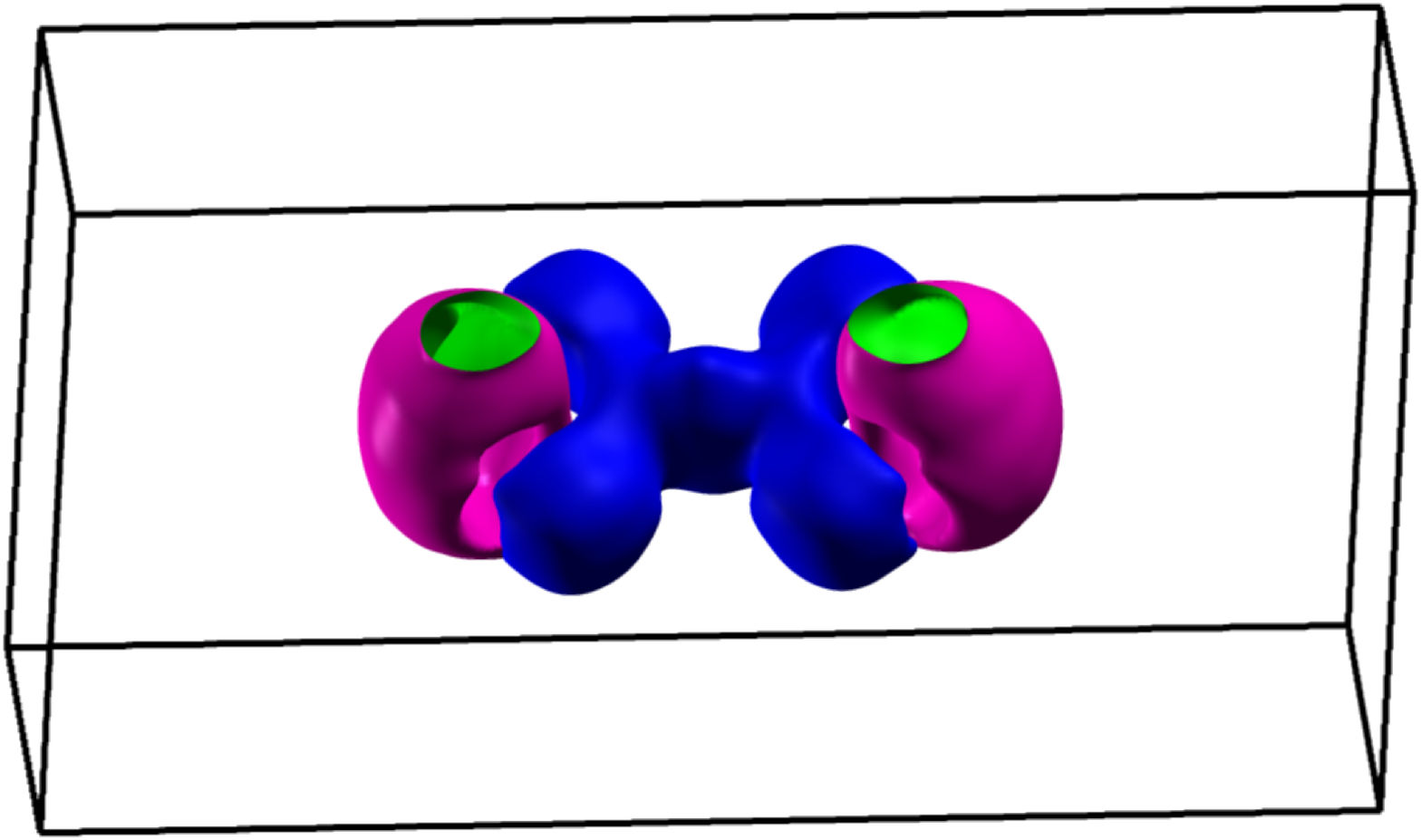}
            \label{fig:U_2_FS}
        }
\hskip 0.01 in
        \subfigure[]{
            \includegraphics[width=1.35 in]{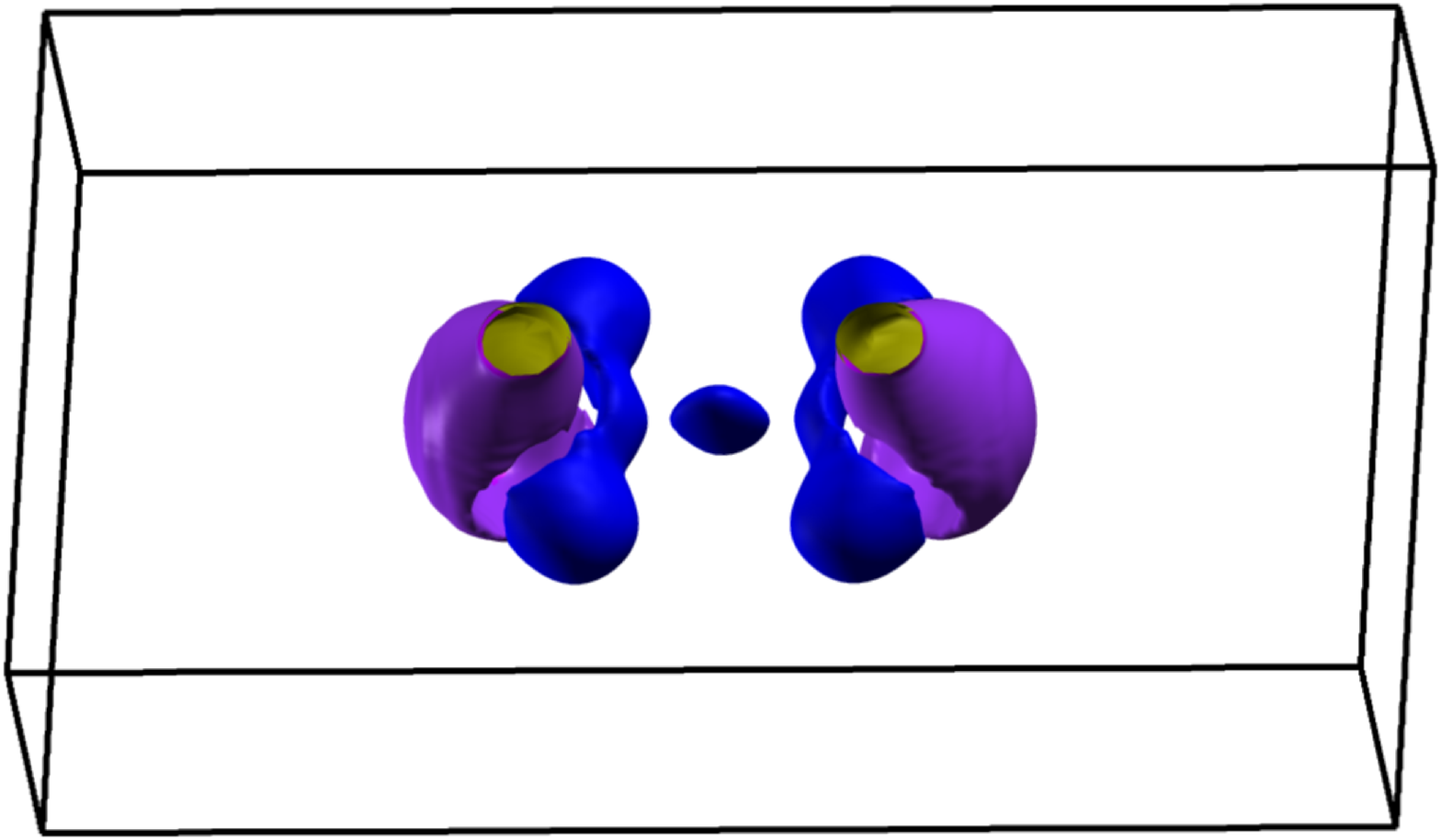}
            \label{fig:U_5_FS}
        }
    \end{center}
        \caption{
(a) Comparison between the band structure of MoTe$_2$ with  U=0 eV (dotted line) and U=3 eV (solid line).
(b),(c) and (d)  Bulk Fermi surface of MoTe$_2$ for U=0, $3$ and $5$ eV respectively.
}
    \label{fig:FS_U}
\end{figure}

In Fig.~\ref{fig:FS_U}, we present the band structure and Fermi surface of MoTe$_2$ obtained for different values 
of a Hubbard U on the Mo-$d$ orbital
(see Fig.~2 of the SM~\onlinecite{Supplementary}  for results using other values of U).
The electron bands around the $Y$ high symmetry point  feel the biggest effect 
as they have the highest proportion of Mo-$d$ orbital.
This is also seen in the FS diagram (Fig.\ref{fig:FS_U} (b) and (c) ) where the smaller hole pockets vanish when U $\ne 0$.
Furthermore, as these electron bands around the $Y$ point are pushed above $E_F$ by the application of U, the electrons in these states occupy 
the lower energy hole bands.
Hence, the hole pockets around the $\Gamma$ point appear to shrink (i.e.,
the hole-bands are pushed down relative to the $E_F$) while they  have negligible Mo-$d$ contributions. 
Interestingly, the direction of movement of the electron and hole bands with respect to the $E_F$ is very similar to the prediction 
of Ref.~\onlinecite{MoTe2Aryal2017} which was done in order to
match the experimental data.
Note that the magnitude of the shift of the electron bands is small compared to the value of U because of the low filling factor of these bands.

Fig.~\ref{fig:FS_U} (b), (c) and (d), illustrates the evolution of the Fermi surface (FS) for U=0, 3 and 5 eV respectively.
Both the electron and hole bands are affected, mostly along the $k_y$ direction 
(i.e., the $Y$ high symmetry direction in the band structure shown in Fig.~\ref{fig:U_3_bands}); the small electron pocket along this direction
disappears and the hole pocket shrinks.
Our FS for non-zero U value is consistent with the HSE calculation of Ref.~\onlinecite{MoTe2HSE_Qi_Nature2016}. 
However, the shape of the FS, in particular that of the hole pocket which change as a function of U, is different from ours.
As a general trend, the ``head'' of the ``star''  hole pocket at $\Gamma$
shrinks by increasing U, whereas 
the ``arms'' stay more or less the same.
For U $ \sim$ 4.5 eV, we find a Lifshitz transition
of the hole pocket where its
central region disconnects from the ``kidney''-like structures as
seen in Fig.~\ref{fig:FS_U} (d).

\begin{figure}[htb]
    \begin{center}
        \subfigure[]{
            \includegraphics[width=1.55 in]{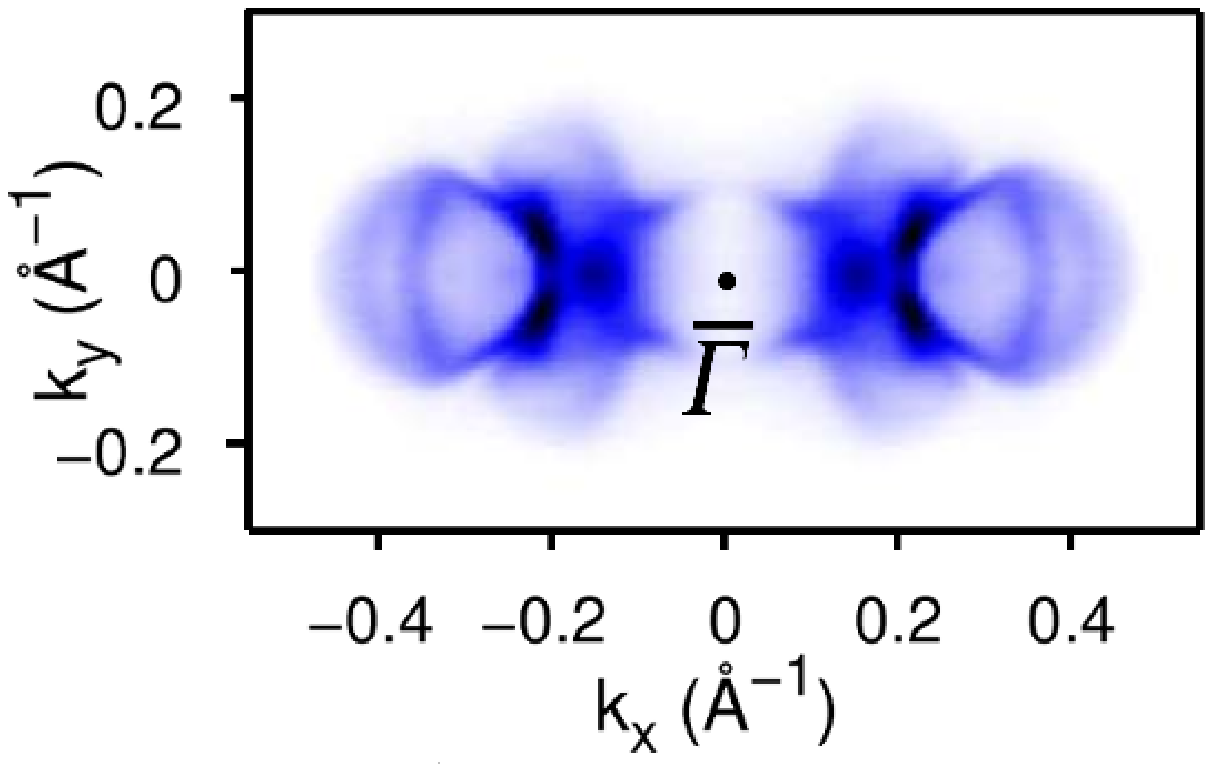}
            \label{fig:arpes}
        }
\hskip 0.01 in
        \subfigure[]{
            \includegraphics[width=1.55 in]{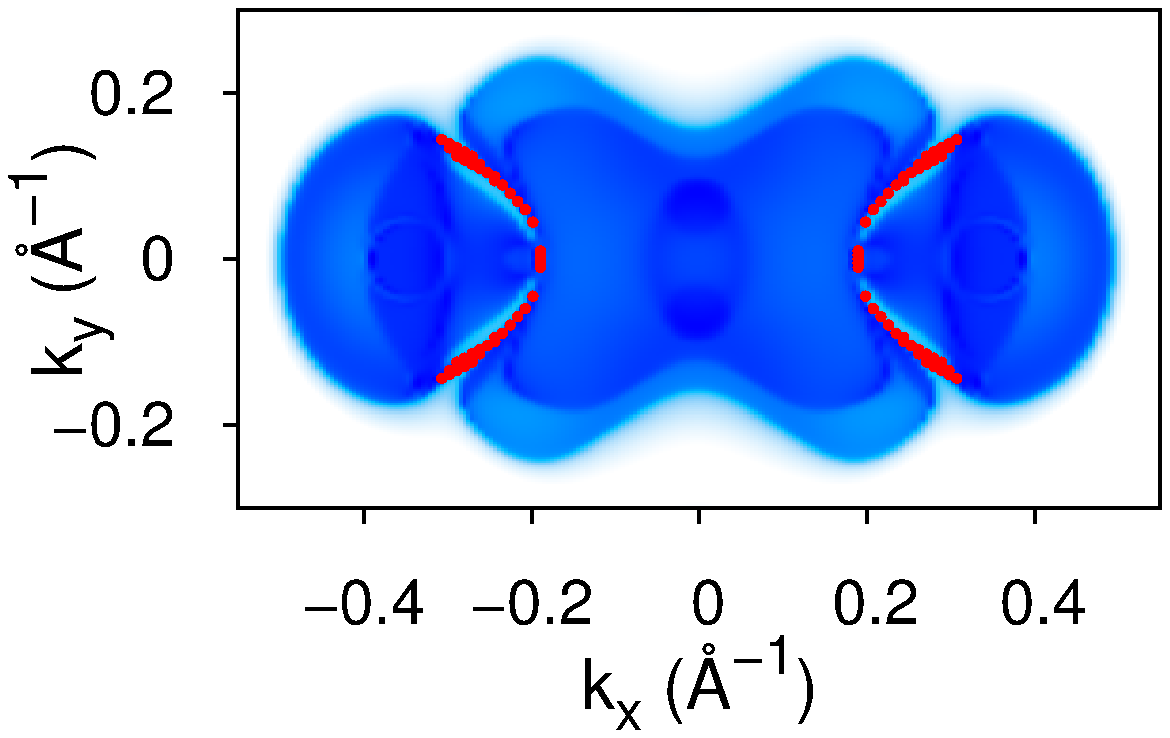}
            \label{fig:DS_FS_DFT}
        }
\vskip 0.01 in
        \subfigure[]{
            \includegraphics[width=1.55 in]{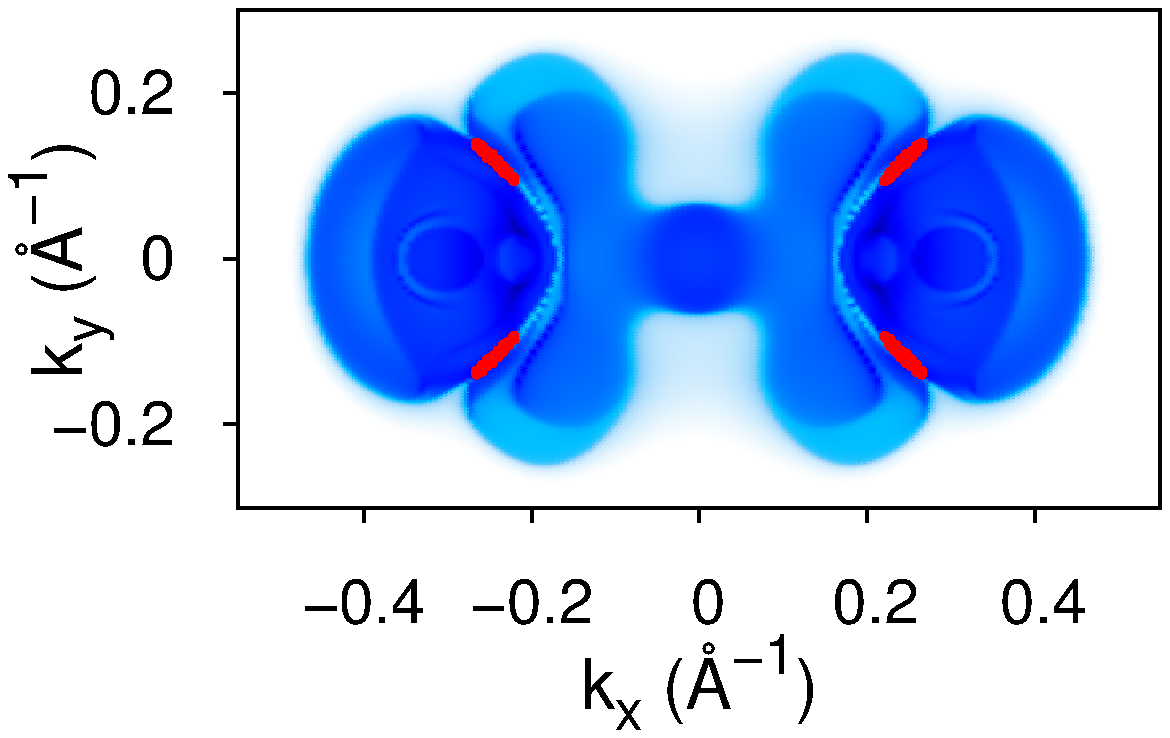}
            \label{fig:DS_FS_U_3}
        }
\hskip 0.01 in
        \subfigure[]{
            \includegraphics[width=1.55 in]{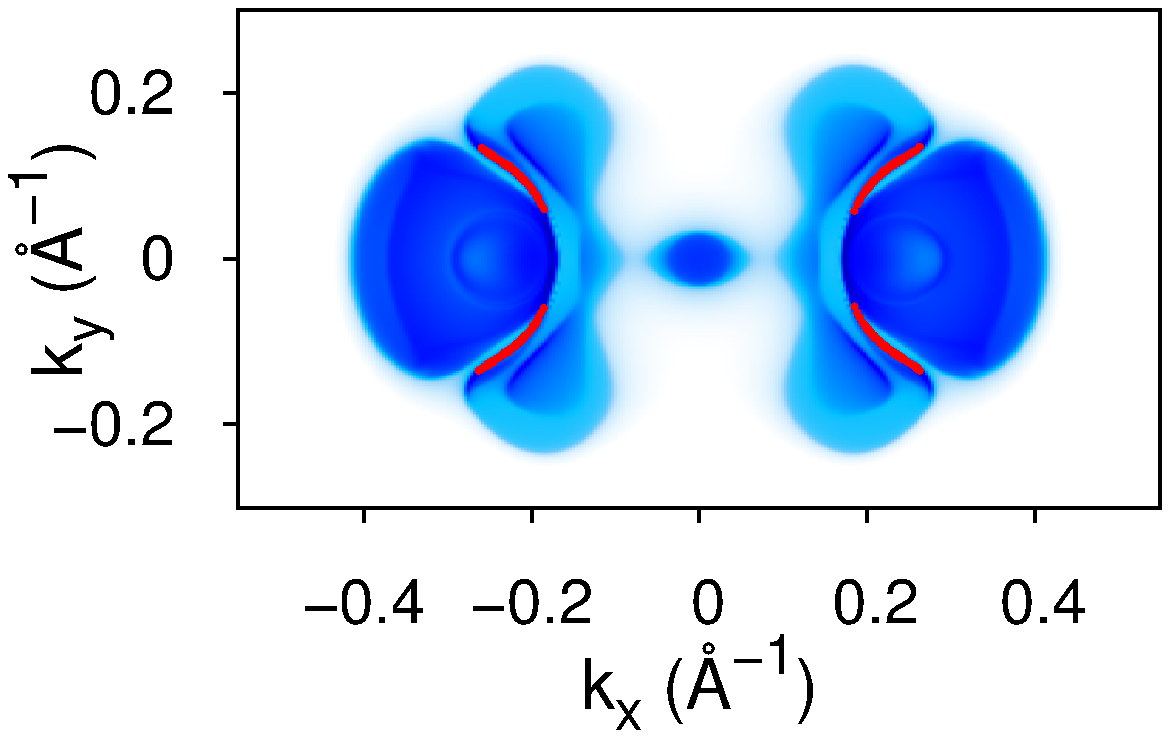}
            \label{fig:DS_FS_U_5}
        }
    \end{center}
        \caption{
 Comparison between ARPES and DFT calculation. 
(a) ARPES measured Fermi surface taken from Ref.~\onlinecite{MoTe2Arpes_jiang_Nature2017}.
 (b,c,d) Calculated projected local density of states at $(0,0,1)$ surface for the slab evaluated at $E=E_F$ for U=0, 3 and 5 eV respectively.
 The surface Fermi-arcs are shown in red and they are in similar positions as found by ARPES.
}
	\label{fig:unpolarized_arpes}
\end{figure}

\begin{figure*}[htb]
    \begin{center}
        \subfigure[]{
            \includegraphics[width=\figw]{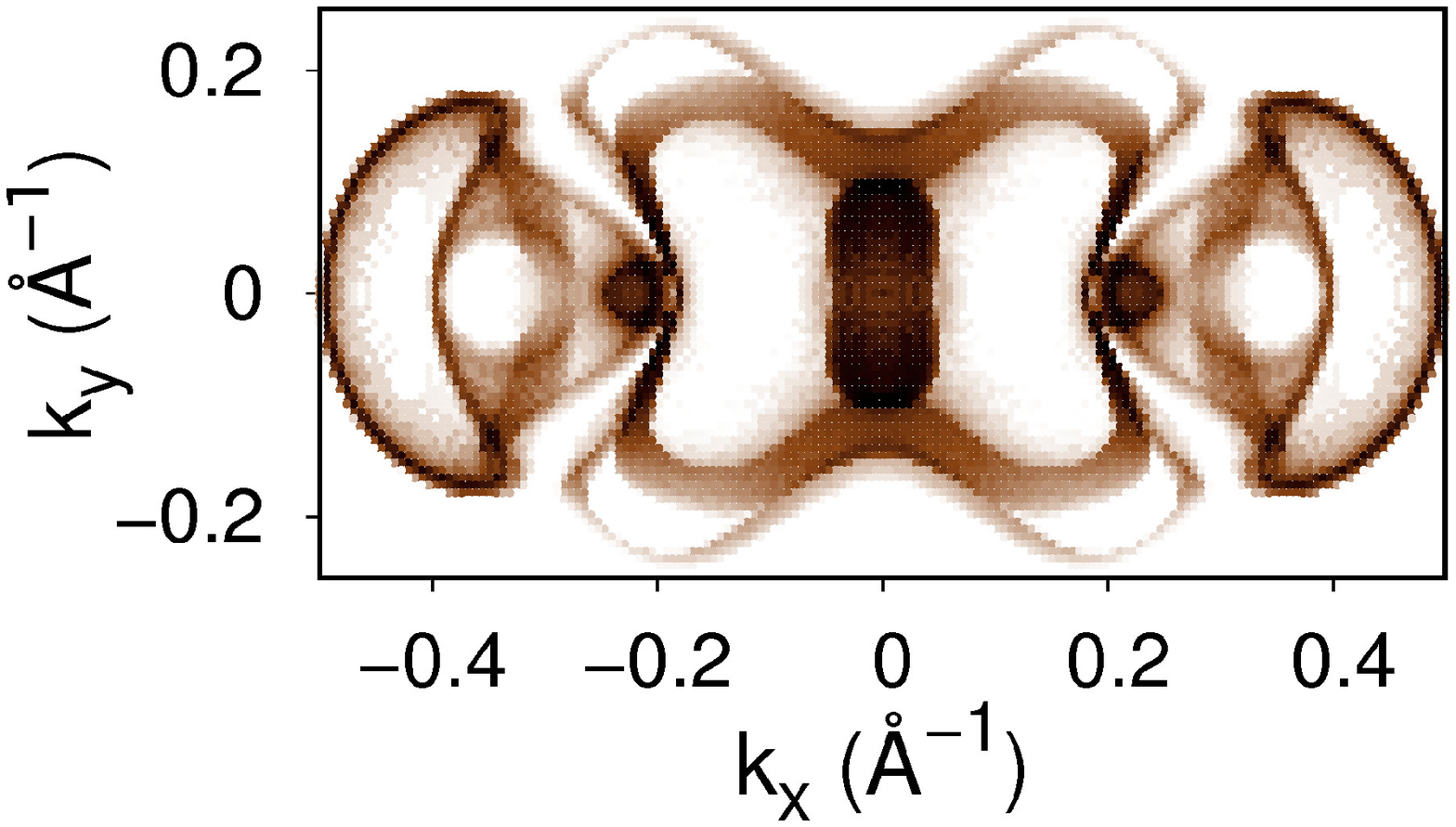}
            \label{fig:proj_FS_DFT_V}
        }
\hskip 0.01 in
        \subfigure[]{
            \includegraphics[width=\figw]{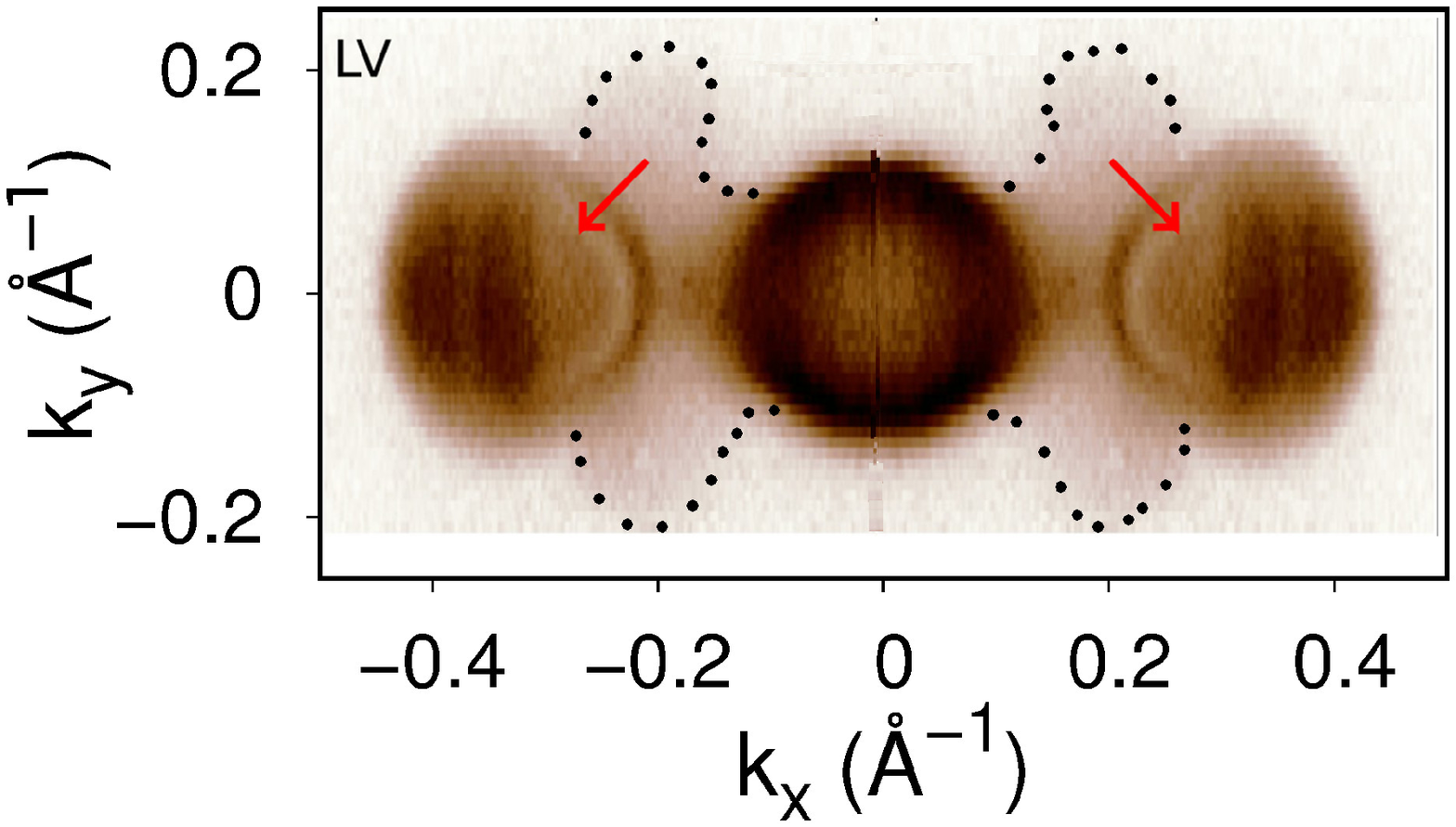}
            \label{fig:arpes_V}
        }
\hskip 0.01 in
        \subfigure[]{
            \includegraphics[width=\figw]{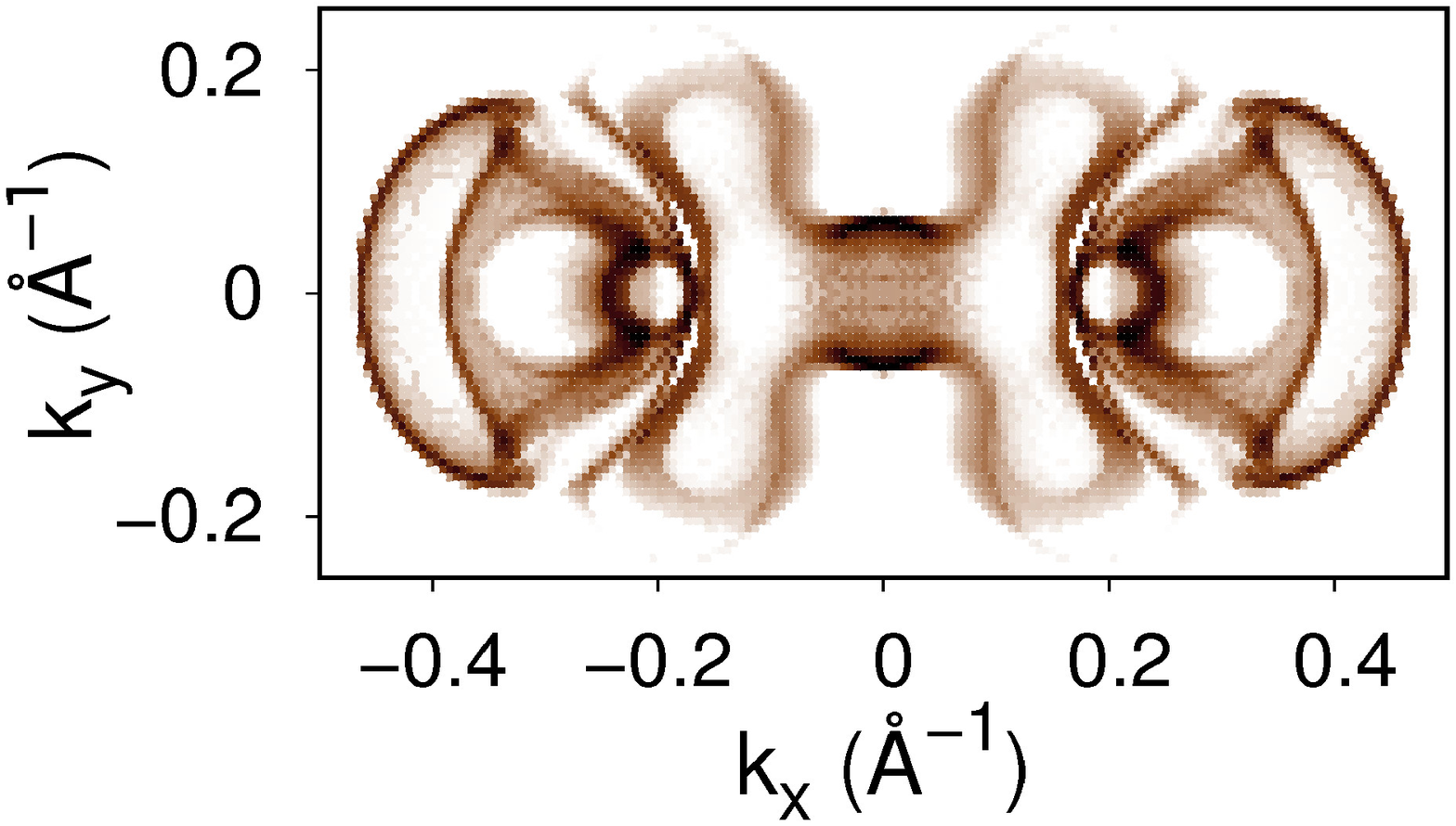}
            \label{fig:proj_FS_U_3_V}
        }
\vskip 0.0 in
        \subfigure[]{
            \includegraphics[width=\figw]{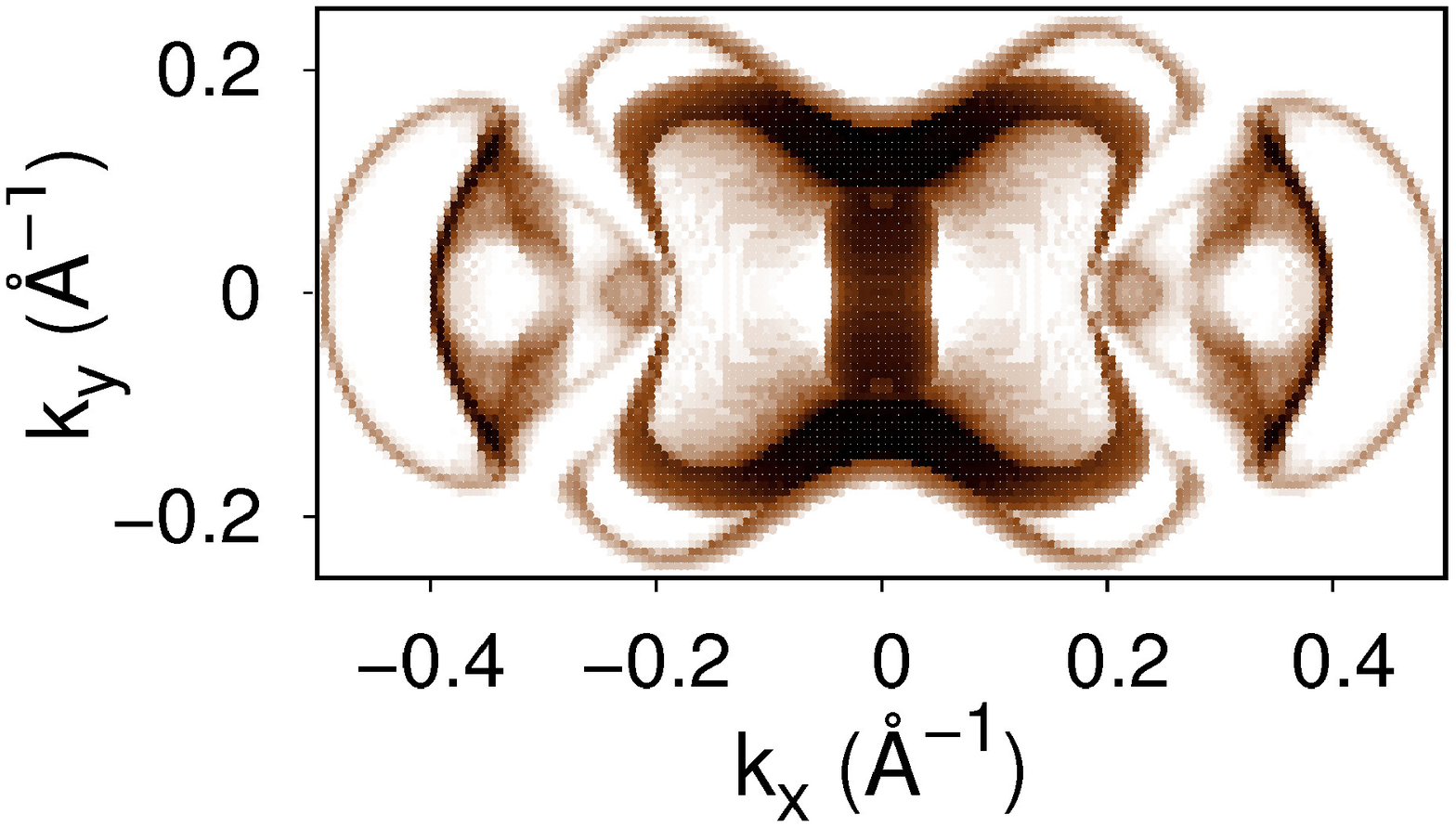}
            \label{fig:proj_FS_DFT_H}
        }
\hskip 0.01 in
        \subfigure[]{
            \includegraphics[width=\figw]{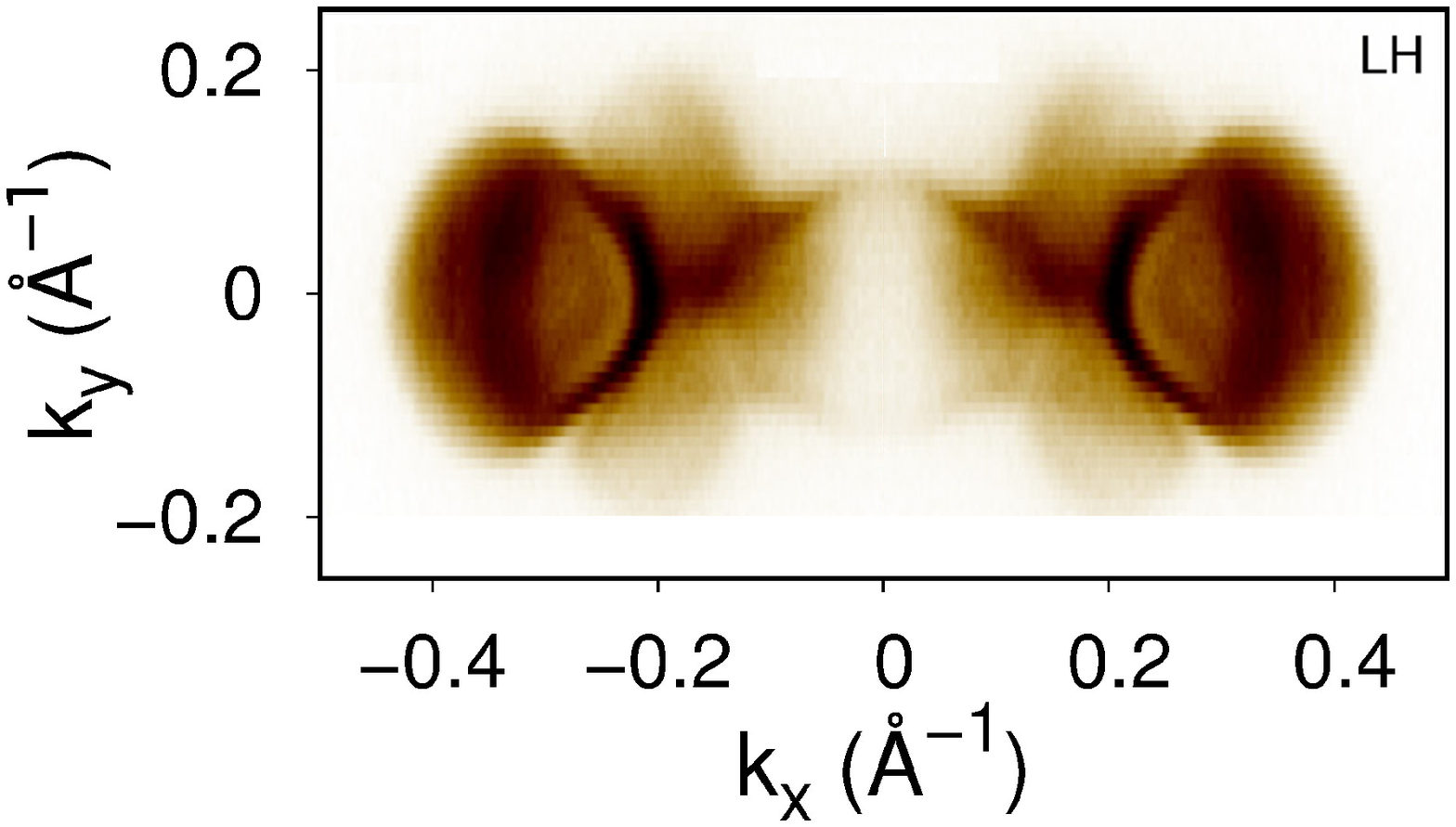}
            \label{fig:arpes_H}
        }
\hskip 0.01 in
        \subfigure[]{
            \includegraphics[width=\figw]{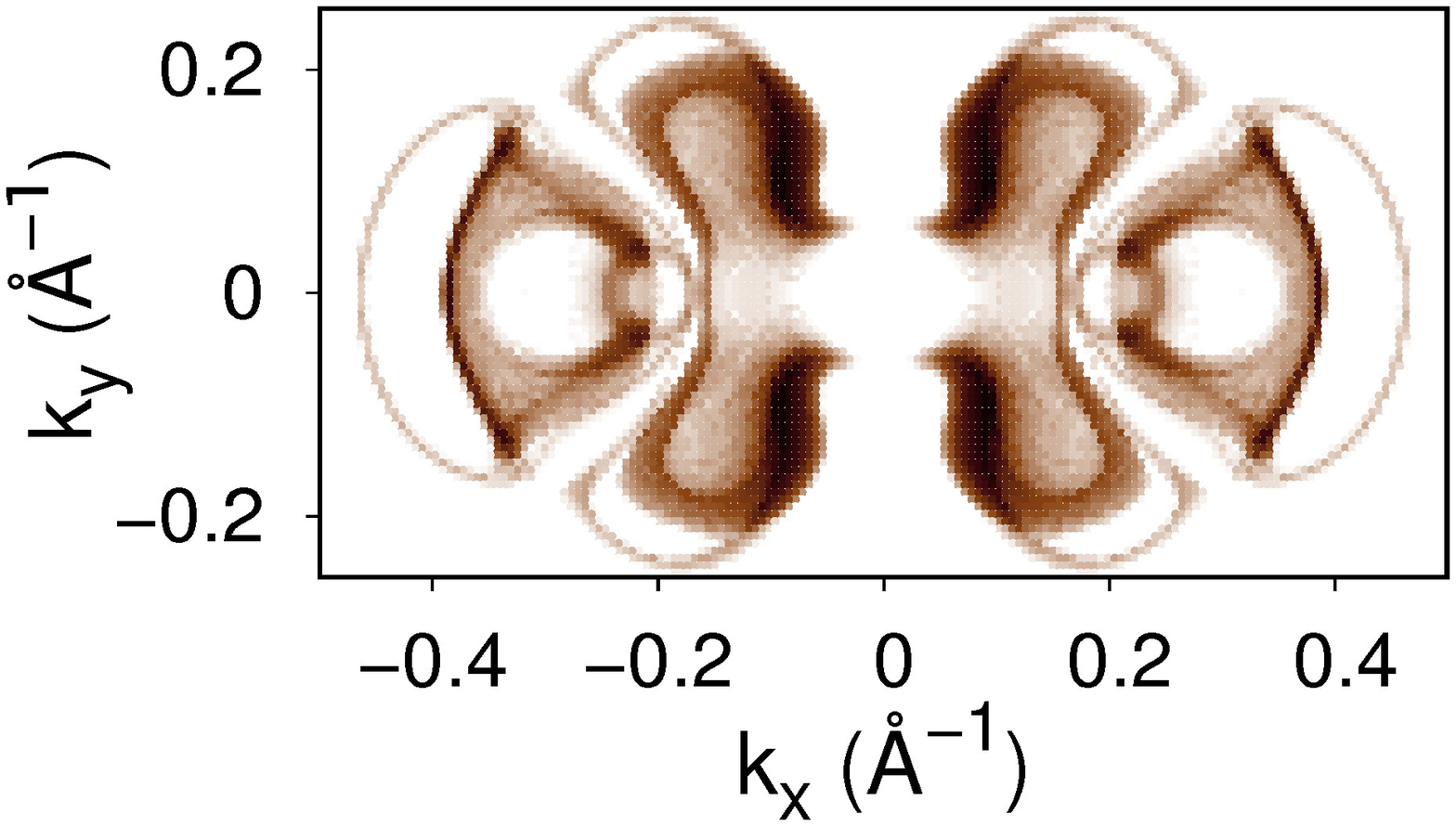}
            \label{fig:proj_FS_U_3_H}
        }
    \end{center}
        \caption{
          Comparison between ARPES and the results of the DFT calculation
          of the polarization dependence. 
          Figs. (a,d) and (c,f) are the calculated bulk FS for U=0  and
          U=3 eV calculations respectively 
          whereas (b) and (e) are ARPES measured FS taken
          from Ref.~\onlinecite{TrivialMoTe2_Crepaldi_Grioni_PRB2017} for the vertical and horizontal polarization of light respectively.
In Figs. (a) and (c) the contribution from only the 
Te-(p$_x$, p$_y$) and Mo-(d$_{x^2-y^2}$, d$_{xy}$ d$_z^2$) orbitals is included
whereas in Figs. (d) and (f), only the contribution for the
Te-(p$_z$) and Mo-(d$_{xz}$, d$_{yz}$) orbitals is included. In Fig.~b we have
added the dotted-line enclosure of the ``kidney''-like features of the
ARPES to guide the eye since for the case of vertical polarization they
are very faint as in our calculations (Fig.~c).
}
	\label{fig:ARPESComparison}
\end{figure*}

In Fig.~\ref{fig:unpolarized_arpes}, we compare the ARPES\cite{MoTe2Arpes_jiang_Nature2017}  results
and the calculated projected  density of states for
different values of U as obtained from a slab calculation.
We can see clearly that DFT overestimates the size of both the hole and electron pockets.
In particular, notice the significant difference between the ARPES and DFT hole pocket around the $\Gamma$ point.
The size of the hole pocket decreases significantly, especially around the $\Gamma$ point when a non-zero U is used.
A good agreement with the ARPES experiment is found for 
U $\sim 3$ eV as seen in Fig.~\ref{fig:DS_FS_U_3}.
As discussed in the previous paragraph, for U $\sim 4.5$ eV a Lifshitz transition  occurs, however, beyond that value of U (Fig.~\ref{fig:DS_FS_U_5}) the central hole pocket  is very small as compared to that seen in the ARPES. 
We note that in the ARPES measurement of Fig.~\ref{fig:arpes}  the central hole pocket is of negligible intensity appearing as white color.
However, different regions of the Fermi surface are visible under different polarizations of light as shown in Ref.~\onlinecite{TrivialMoTe2_Crepaldi_Grioni_PRB2017}.
Next, by calculating the ARPES light-polarization dependence
we show that while
it cannot be captured by the DFT calculation,  the DFT+U calculation yields
a good
qualitative account.

The bands that cross the Fermi surface have different distribution of
the $p$ and $d$ orbitals which are 
visible under different directions of polarization of light.
As shown in Sec.~III of the SM\cite{Supplementary}, the
matrix elements of
the operator $\vec \epsilon \cdot \nabla$ (where $\vec \epsilon$ is the
direction of the light polarization) which enters in 
ARPES intensity follows the following rules
for the different orbital character of the components of each band.
(a) When $\vec \epsilon || \hat z$, from the $p$-orbitals
  only the $p_z$ character contributes
  to the intensity. From the $d$-orbitals, only the $d_{xz}$ and $d_{yz}$
    contribute.
    (b) When $\vec \epsilon \perp \hat z$, from the $p$-orbitals
  only the $p_x$ and $p_y$ character contribute
  to the intensity. From the $d$-orbitals, only the $d_{xy}$ and $d_{x^2-y^2}$
  and $d_{z^2}$
  contribute.

In Fig.~\ref{fig:ARPESComparison} we present the results of our calculation where we only include
the non-zero contributions for a given direction of light polarization.
We also compare these results obtained with U=0 (DFT) and U=3 eV to the
ARPES measurements\cite{TrivialMoTe2_Crepaldi_Grioni_PRB2017}.
Figs.~\ref{fig:proj_FS_DFT_V},\ref{fig:proj_FS_DFT_H} and
Figs.~\ref{fig:proj_FS_U_3_V},\ref{fig:proj_FS_U_3_H} are the calculated bulk FS for U=0  and
          U=3 eV calculations respectively, 
          whereas Figs.~\ref{fig:arpes_V},\ref{fig:arpes_H} are the ARPES measured FS for the vertical and horizontal polarization of light respectively.
          The main feature of the results is that the part of the hole pocket
          which encircles the $\Gamma$ point has strong polarization dependence
          which is captured by our calculation for U=3 eV, while the
          DFT calculation fails to do so. In addition, the ``kidney''-like parts of the hole pocket are faint in the case of vertical polarization and more intense in the case of horizontal polarization, a feature which is also present
          in the U=3 eV calculation and is absent in the DFT calculation. We consider these features which can be captured only by the
          DFT+U calculation a ``smoking gun'' that the DFT+U calculation gives a much better picture of what happens near the
          FS of $\gamma$-MoTe$_2$.

Direct comparison of the calculated angular dependence with the experimentally measured QO data did not turn out to be straightforward 
because we found that the calculation shows significant difference between the spin-orbit (SO) split bands. 
This is in contrary to the experimentally observed QO FFT spectra (see Fig.~4 (b) of Ref.~\onlinecite{MoTe2Aryal2017}) where the twin peaks with similar angular dependence, identified as
SO partners, have much smaller SO splittings.
Overestimation of SOC strengths in DFT calculations, thereby giving inaccurate
values of experimental observables, is
not a new finding by itself and has been discussed
in previous works\cite{RoleSOI_Radzynski2011}.
Hence, we averaged over the angular dependence of the SO partners which 
is justified because of their similar angular dependence
(see Fig.~4 of the SM~\cite{Supplementary}
for comparison  without averaging).

\begin{figure}[htb]
    \begin{center}
	\vskip 0.2 in
            \includegraphics[width=3.4 in]{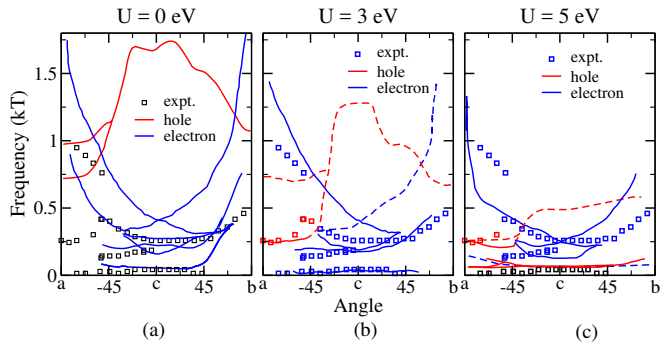}
    \end{center}
        \caption{
Comparison between the experimental angular dependence and the calculated angular dependence 
for (a) DFT (b) U=3 eV (c) U=5 eV.
In Figs.~(b) and (c), the calculated orbits that do not correspond to the experimental ones are drawn with dashed lines.
}
\label{fig:angular_dep_exp_theory}
\end{figure}

In Ref.~\onlinecite{MoTe2Aryal2017}, it was shown that the DFT calculated angular dependence of the Fermi surface does not agree with 
the experimentally measured QO frequencies. In particular, a
larger than 1000 T QO frequency along the $c$-axis is absent
and this is also true for the QO frequencies reported
in Ref.~\onlinecite{MoTe2HSE_Qi_Nature2016}.
In Fig.~\ref{fig:angular_dep_exp_theory} the
angular dependence of the orbits measured by QO experiments is
compared to our results using U=0, 3 and 5 eV.
In Fig.~\ref{fig:angular_dep_exp_theory}(a) and Fig.~\ref{fig:angular_dep_exp_theory}(b), both the spin-orbit partners are averaged whereas in
Fig.~\ref{fig:angular_dep_exp_theory}(c), only the electron partners are averaged, as the smaller of the hole pockets further disintegrates into smaller pieces and hence, they have different angular dependence. 
Though the size of both the electron and hole pockets
decrease as a function of U, the hole pockets are the ones most affected. 
When the U value is increased in order to obtain the
Lifshitz transition,  the large hole pocket of size 
$\approx 1500$ T collapses to two small sized pockets of approximate sizes 
$ 500 $ T and  $100$ T.
(see also Fig.~3 (b) of the SM~\cite{Supplementary} for other values of U).
This effect is dramatic and can provide a plausible answer for the absence of large sized hole pockets in the QO experiments.
However, as can be inferred from Fig.~\ref{fig:ARPESComparison}, this is contrary to the observation of the  ARPES experiments.
On the other hand, if we adopt a moderate value of U=3 eV, it can account for
 most of the Quantum Oscillation frequencies 
 (including the hole orbit along the $a$-axis) and at the same time can
 explain the
 ARPES experiments including the polarization dependence of the
 intensity.
 The large hole orbit of size greater than 1000 T not seen in QO experiments
 along 
 the $c$-axis can not be accounted with this value of U; however, it is not
unusual in a QO experiment to miss an orbit along
a certain direction, 
especially when the mass of the carrier is very anisotropic\cite{WP2_Schonemann_Aryal_PRB_2017,MAl3_KWChen_PRL2018}. The fact, however, that $\gamma$-MoTe$_2$
is near the Lifshitz transition, which in practical terms means
just 25 meV (we find that this corresponds to less than $\sim$  2 \% per
chemical formula electron doping) away from the transition, it is expected
to have a significant effect in transport phenomena. The reason
is that the size of the large hole pocket near $\Gamma$ changes rapidly by
external causes.

\label{RobustnessWeyl}
\begin{figure}[htb]
    \begin{center}
\vskip 0.2 in
            \includegraphics[width=3.0 in]{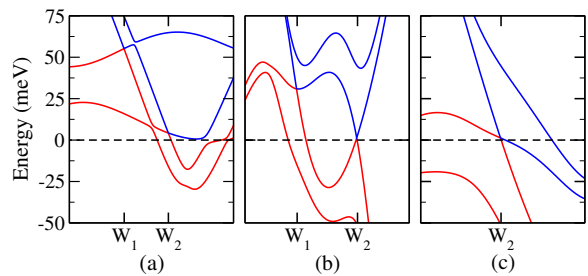}
    \end{center}
        \caption{
Band structure along the Weyl points W$_1$ and W$_2$ for  (a) U = 0 eV, (b) U = 3 eV  and in (c) 
for U = 3 eV along a direction perpendicular to the W$_1$W$_2$ line which passes through the W$_2$ point and lies on the XY plane.
}
	\label{fig:WeylPoint_U_0_3}
\end{figure}
As noted by other authors\cite{Wang_Bernevig_2016,MoTe2YanPRB2015,CorrelationsWTe2_Di_Panaccione_PRL_2017,Bruno_Baumberger_2016}, the 
exact position and even the presence of WPs is very sensitive to 
the slightest change in the lattice parameters and also in the different implementations of DFT.
Therefore, first we identify the position of the WPs for
MoTe$_2$ by looking at the Berry curvature vector plot. 
(See Fig.~5 (a) and (b) of the SM\cite{Supplementary} which illustrates the location of the source and sink of 
Berry curvature.)
In total, there are 4 pairs of WPs, each on one quadrant, all of which are located  on the $k_z=0$ plane and are related by the crystal symmetry. 
The band structure along these WPs is shown in Fig.~\ref{fig:WeylPoint_U_0_3} (a)
which is similar to the one shown in Ref.\onlinecite{MoTe2YanPRB2015}.

We also examined how the WPs evolve as a function of U.
We found that the number of the WPs and their position in the BZ change
with increasing U in a very non-trivial way.
For example, for U=0.75 eV, we find 12 pairs of 
WPs out of which only 4 pairs are located on the $k_z=0$ plane.
For U=2 eV and U=4 eV, there are respectively 8 and 6 pairs of WPs,
none of which are located on the $k_z=0$ plane.
Also, we found that the source and sinks of the WPs 
move farther from one-another as U increases.
In Fig.\ref{fig:WeylPoint_U_0_3}, we present the band structure along the WPs for U=3 eV along the direction W$_1$W$_2$
(Fig.~\ref{fig:WeylPoint_U_0_3}(b))
and along the perpendicular direction (Fig.~\ref{fig:WeylPoint_U_0_3}(c)) where it is shown that one pair of surviving WPs is
very close to the $E_F$. Notice that the WP, which is very close to
the FS, seems to be a type-I
along the  W$_1$W$_2$ direction; however,  when we calculated the band
dispersion along the direction perpendicular to W$_1$W$_2$ (Fig.~\ref{fig:WeylPoint_U_0_3}(c)), it becomes clear that its character is of type-II.
In Sec.~IV of the SM~\cite{Supplementary}, we introduce a simple model for
an inversion symmetry breaking quasi-2D system and demonstrate how the
WPs can be destroyed or created by the presence of the Hubbard U.
Using this model  we find that the WPs move in $k$-space with the application of U until they  
annihilate. 

In summary, we
found that a modest value of U (close to 3 eV) gives good qualitative agreement
with both the 
QO experiments and the ARPES measured Fermi surface. The ``smoking gun'' that
our calculation provides a much more accurate interpretation of the FS of
$\gamma$-MoTe$_2$ is that it can explain the polarization dependence of
ARPES, while the DFT calculation fails.
We find a Lifshitz transition for U $\sim 4.5$ eV and, because of the
more accurate account of the experimental results by using a not too different
value of U,
this suggests that $\gamma$-MoTe$_2$ could be in the close vicinity to this 
transition produced by the effects of Coulomb correlations.
Experimentally, the Lifshitz transition can be probed by a
small amount of doping which can be introduced either chemically or by making
a device with a back-gate potential which can effectively introduce charge. 
We find that the number of WPs and their position in the BZ
change non-linearly with U.
Interestingly, the WPs closest to the Fermi level survive for the U value which
accurately describes both ARPES and QO experiments
and this is a very positive result of the present
work regarding a controversial issue.
\section{Acknowledgments}

We wish to thank L. Balicas and K.-W. Chen for useful discussions.
This work was supported in part by the U.S. National High Magnetic Field
Laboratory, which is
funded by NSF DMR-1157490 and the State of Florida.


\begin{thebibliography}{28}%
\makeatletter
\providecommand \@ifxundefined [1]{%
 \@ifx{#1\undefined}
}%
\providecommand \@ifnum [1]{%
 \ifnum #1\expandafter \@firstoftwo
 \else \expandafter \@secondoftwo
 \fi
}%
\providecommand \@ifx [1]{%
 \ifx #1\expandafter \@firstoftwo
 \else \expandafter \@secondoftwo
 \fi
}%
\providecommand \natexlab [1]{#1}%
\providecommand \enquote  [1]{``#1''}%
\providecommand \bibnamefont  [1]{#1}%
\providecommand \bibfnamefont [1]{#1}%
\providecommand \citenamefont [1]{#1}%
\providecommand \href@noop [0]{\@secondoftwo}%
\providecommand \href [0]{\begingroup \@sanitize@url \@href}%
\providecommand \@href[1]{\@@startlink{#1}\@@href}%
\providecommand \@@href[1]{\endgroup#1\@@endlink}%
\providecommand \@sanitize@url [0]{\catcode `\\12\catcode `\$12\catcode
  `\&12\catcode `\#12\catcode `\^12\catcode `\_12\catcode `\%12\relax}%
\providecommand \@@startlink[1]{}%
\providecommand \@@endlink[0]{}%
\providecommand \url  [0]{\begingroup\@sanitize@url \@url }%
\providecommand \@url [1]{\endgroup\@href {#1}{\urlprefix }}%
\providecommand \urlprefix  [0]{URL }%
\providecommand \Eprint [0]{\href }%
\providecommand \doibase [0]{http://dx.doi.org/}%
\providecommand \selectlanguage [0]{\@gobble}%
\providecommand \bibinfo  [0]{\@secondoftwo}%
\providecommand \bibfield  [0]{\@secondoftwo}%
\providecommand \translation [1]{[#1]}%
\providecommand \BibitemOpen [0]{}%
\providecommand \bibitemStop [0]{}%
\providecommand \bibitemNoStop [0]{.\EOS\space}%
\providecommand \EOS [0]{\spacefactor3000\relax}%
\providecommand \BibitemShut  [1]{\csname bibitem#1\endcsname}%
\let\auto@bib@innerbib\@empty
\bibitem [{\citenamefont {Wang}\ \emph {et~al.}(2012)\citenamefont {Wang},
  \citenamefont {Kalantar-Zadeh}, \citenamefont {Kis}, \citenamefont
  {Coleman},\ and\ \citenamefont {Strano}}]{TMDsWangnature2012}%
  \BibitemOpen
  \bibfield  {author} {\bibinfo {author} {\bibfnamefont {Q.~H.}\ \bibnamefont
  {Wang}}, \bibinfo {author} {\bibfnamefont {K.}~\bibnamefont
  {Kalantar-Zadeh}}, \bibinfo {author} {\bibfnamefont {A.}~\bibnamefont {Kis}},
  \bibinfo {author} {\bibfnamefont {J.~N.}\ \bibnamefont {Coleman}}, \ and\
  \bibinfo {author} {\bibfnamefont {M.~S.}\ \bibnamefont {Strano}},\ }\href
  {http://dx.doi.org/10.1038/nnano.2012.193} {\bibfield  {journal} {\bibinfo
  {journal} {Nat. Nanotechnol.}\ }\textbf {\bibinfo {volume} {7}},\ \bibinfo
  {pages} {699} (\bibinfo {year} {2012})}\BibitemShut {NoStop}%
\bibitem [{\citenamefont {Wilson}\ and\ \citenamefont
  {Yoffe}(1969)}]{TMDsWilsonAiP1969}%
  \BibitemOpen
  \bibfield  {author} {\bibinfo {author} {\bibfnamefont {J.}~\bibnamefont
  {Wilson}}\ and\ \bibinfo {author} {\bibfnamefont {A.}~\bibnamefont {Yoffe}},\
  }\href {\doibase 10.1080/00018736900101307} {\bibfield  {journal} {\bibinfo
  {journal} {Adv. Phys.}\ }\textbf {\bibinfo {volume} {18}},\ \bibinfo {pages}
  {193} (\bibinfo {year} {1969})}\BibitemShut {NoStop}%
\bibitem [{\citenamefont {Mattheiss}(1973)}]{TMDsMattheissPRB1973}%
  \BibitemOpen
  \bibfield  {author} {\bibinfo {author} {\bibfnamefont {L.~F.}\ \bibnamefont
  {Mattheiss}},\ }\href {\doibase 10.1103/PhysRevB.8.3719} {\bibfield
  {journal} {\bibinfo  {journal} {Phys. Rev. B}\ }\textbf {\bibinfo {volume}
  {8}},\ \bibinfo {pages} {3719} (\bibinfo {year} {1973})}\BibitemShut
  {NoStop}%
\bibitem [{\citenamefont {Wang}\ \emph
  {et~al.}(2016{\natexlab{a}})\citenamefont {Wang}, \citenamefont {Gresch},
  \citenamefont {Soluyanov}, \citenamefont {Xie}, \citenamefont {Kushwaha},
  \citenamefont {Dai}, \citenamefont {Troyer}, \citenamefont {Cava},\ and\
  \citenamefont {Bernevig}}]{Wang_Bernevig_2016}%
  \BibitemOpen
  \bibfield  {author} {\bibinfo {author} {\bibfnamefont {Z.}~\bibnamefont
  {Wang}}, \bibinfo {author} {\bibfnamefont {D.}~\bibnamefont {Gresch}},
  \bibinfo {author} {\bibfnamefont {A.~A.}\ \bibnamefont {Soluyanov}}, \bibinfo
  {author} {\bibfnamefont {W.}~\bibnamefont {Xie}}, \bibinfo {author}
  {\bibfnamefont {S.}~\bibnamefont {Kushwaha}}, \bibinfo {author}
  {\bibfnamefont {X.}~\bibnamefont {Dai}}, \bibinfo {author} {\bibfnamefont
  {M.}~\bibnamefont {Troyer}}, \bibinfo {author} {\bibfnamefont {R.~J.}\
  \bibnamefont {Cava}}, \ and\ \bibinfo {author} {\bibfnamefont {B.~A.}\
  \bibnamefont {Bernevig}},\ }\href {\doibase 10.1103/PhysRevLett.117.056805}
  {\bibfield  {journal} {\bibinfo  {journal} {Phys. Rev. Lett.}\ }\textbf
  {\bibinfo {volume} {117}},\ \bibinfo {pages} {056805} (\bibinfo {year}
  {2016}{\natexlab{a}})}\BibitemShut {NoStop}%
\bibitem [{\citenamefont {Soluyanov}\ \emph {et~al.}(2015)\citenamefont
  {Soluyanov}, \citenamefont {Gresch}, \citenamefont {Wang}, \citenamefont
  {Wu}, \citenamefont {Troyer}, \citenamefont {Dai},\ and\ \citenamefont
  {Bernevig}}]{TypeIISoluyanov2015}%
  \BibitemOpen
  \bibfield  {author} {\bibinfo {author} {\bibfnamefont {A.~A.}\ \bibnamefont
  {Soluyanov}}, \bibinfo {author} {\bibfnamefont {D.}~\bibnamefont {Gresch}},
  \bibinfo {author} {\bibfnamefont {Z.}~\bibnamefont {Wang}}, \bibinfo {author}
  {\bibfnamefont {Q.}~\bibnamefont {Wu}}, \bibinfo {author} {\bibfnamefont
  {M.}~\bibnamefont {Troyer}}, \bibinfo {author} {\bibfnamefont
  {X.}~\bibnamefont {Dai}}, \ and\ \bibinfo {author} {\bibfnamefont {B.~A.}\
  \bibnamefont {Bernevig}},\ }\href {http://dx.doi.org/10.1038/nature15768}
  {\bibfield  {journal} {\bibinfo  {journal} {Nature}\ }\textbf {\bibinfo
  {volume} {527}},\ \bibinfo {pages} {495} (\bibinfo {year}
  {2015})}\BibitemShut {NoStop}%
\bibitem [{\citenamefont {Chang}\ \emph {et~al.}(2016)\citenamefont {Chang},
  \citenamefont {Xu}, \citenamefont {Chang}, \citenamefont {Lee}, \citenamefont
  {Huang} \emph {et~al.}}]{Prediction_ChangNature_2016}%
  \BibitemOpen
  \bibfield  {author} {\bibinfo {author} {\bibfnamefont {T.-R.}\ \bibnamefont
  {Chang}}, \bibinfo {author} {\bibfnamefont {S.-Y.}\ \bibnamefont {Xu}},
  \bibinfo {author} {\bibfnamefont {G.}~\bibnamefont {Chang}}, \bibinfo
  {author} {\bibfnamefont {C.-C.}\ \bibnamefont {Lee}}, \bibinfo {author}
  {\bibfnamefont {S.-M.}\ \bibnamefont {Huang}},  \emph {et~al.},\ }\href
  {http://dx.doi.org/10.1038/ncomms10639} {\bibfield  {journal} {\bibinfo
  {journal} {Nat. Commun.}\ }\textbf {\bibinfo {volume} {7}},\ \bibinfo {pages}
  {10639} (\bibinfo {year} {2016})}\BibitemShut {NoStop}%
\bibitem [{\citenamefont {Xu}\ \emph {et~al.}(2015{\natexlab{a}})\citenamefont
  {Xu}, \citenamefont {Belopolski}, \citenamefont {Alidoust}, \citenamefont
  {Neupane}, \citenamefont {Bian} \emph
  {et~al.}}]{FermiarcsTaAs_XuScience2015}%
  \BibitemOpen
  \bibfield  {author} {\bibinfo {author} {\bibfnamefont {S.-Y.}\ \bibnamefont
  {Xu}}, \bibinfo {author} {\bibfnamefont {I.}~\bibnamefont {Belopolski}},
  \bibinfo {author} {\bibfnamefont {N.}~\bibnamefont {Alidoust}}, \bibinfo
  {author} {\bibfnamefont {M.}~\bibnamefont {Neupane}}, \bibinfo {author}
  {\bibfnamefont {G.}~\bibnamefont {Bian}},  \emph {et~al.},\ }\href {\doibase
  10.1126/science.aaa9297} {\bibfield  {journal} {\bibinfo  {journal}
  {Science}\ }\textbf {\bibinfo {volume} {349}},\ \bibinfo {pages} {613}
  (\bibinfo {year} {2015}{\natexlab{a}})}\BibitemShut {NoStop}%
\bibitem [{\citenamefont {Huang}\ \emph
  {et~al.}(2015{\natexlab{a}})\citenamefont {Huang}, \citenamefont {Xu},
  \citenamefont {Belopolski}, \citenamefont {Lee}, \citenamefont {Chang} \emph
  {et~al.}}]{FermiarcsTaAs_HuangNature2015}%
  \BibitemOpen
  \bibfield  {author} {\bibinfo {author} {\bibfnamefont {S.-M.}\ \bibnamefont
  {Huang}}, \bibinfo {author} {\bibfnamefont {S.-Y.}\ \bibnamefont {Xu}},
  \bibinfo {author} {\bibfnamefont {I.}~\bibnamefont {Belopolski}}, \bibinfo
  {author} {\bibfnamefont {C.-C.}\ \bibnamefont {Lee}}, \bibinfo {author}
  {\bibfnamefont {G.}~\bibnamefont {Chang}},  \emph {et~al.},\ }\href
  {http://dx.doi.org/10.1038/ncomms8373} {\bibfield  {journal} {\bibinfo
  {journal} {Nat. Commun.}\ }\textbf {\bibinfo {volume} {6}},\ \bibinfo {pages}
  {7373} (\bibinfo {year} {2015}{\natexlab{a}})}\BibitemShut {NoStop}%
\bibitem [{\citenamefont {Xu}\ \emph {et~al.}(2015{\natexlab{b}})\citenamefont
  {Xu}, \citenamefont {Alidoust}, \citenamefont {Belopolski}, \citenamefont
  {Yuan}, \citenamefont {Bian}, \citenamefont {Chang} \emph
  {et~al.}}]{FermiarcsNbAs_XuNature2015}%
  \BibitemOpen
  \bibfield  {author} {\bibinfo {author} {\bibfnamefont {S.-Y.}\ \bibnamefont
  {Xu}}, \bibinfo {author} {\bibfnamefont {N.}~\bibnamefont {Alidoust}},
  \bibinfo {author} {\bibfnamefont {I.}~\bibnamefont {Belopolski}}, \bibinfo
  {author} {\bibfnamefont {Z.}~\bibnamefont {Yuan}}, \bibinfo {author}
  {\bibfnamefont {G.}~\bibnamefont {Bian}}, \bibinfo {author} {\bibnamefont
  {Chang}},  \emph {et~al.},\ }\href {http://dx.doi.org/10.1038/nphys3437}
  {\bibfield  {journal} {\bibinfo  {journal} {Nat. Phys.}\ }\textbf {\bibinfo
  {volume} {11}},\ \bibinfo {pages} {748} (\bibinfo {year}
  {2015}{\natexlab{b}})}\BibitemShut {NoStop}%
\bibitem [{\citenamefont {Armitage}\ \emph {et~al.}(2018)\citenamefont
  {Armitage}, \citenamefont {Mele},\ and\ \citenamefont
  {Vishwanath}}]{Armitage_Vishwanath_RMP2018}%
  \BibitemOpen
  \bibfield  {author} {\bibinfo {author} {\bibfnamefont {N.~P.}\ \bibnamefont
  {Armitage}}, \bibinfo {author} {\bibfnamefont {E.~J.}\ \bibnamefont {Mele}},
  \ and\ \bibinfo {author} {\bibfnamefont {A.}~\bibnamefont {Vishwanath}},\
  }\href {\doibase 10.1103/RevModPhys.90.015001} {\bibfield  {journal}
  {\bibinfo  {journal} {Rev. Mod. Phys.}\ }\textbf {\bibinfo {volume} {90}},\
  \bibinfo {pages} {015001} (\bibinfo {year} {2018})}\BibitemShut {NoStop}%
\bibitem [{\citenamefont {Wang}\ \emph
  {et~al.}(2016{\natexlab{b}})\citenamefont {Wang}, \citenamefont {Zheng},
  \citenamefont {Shen}, \citenamefont {Lu}, \citenamefont {Fang}, \citenamefont
  {Sheng}, \citenamefont {Zhou}, \citenamefont {Yang}, \citenamefont {Li},
  \citenamefont {Feng},\ and\ \citenamefont
  {Xu}}]{UltraHighMobilityNbPWangPRB2016}%
  \BibitemOpen
  \bibfield  {author} {\bibinfo {author} {\bibfnamefont {Z.}~\bibnamefont
  {Wang}}, \bibinfo {author} {\bibfnamefont {Y.}~\bibnamefont {Zheng}},
  \bibinfo {author} {\bibfnamefont {Z.}~\bibnamefont {Shen}}, \bibinfo {author}
  {\bibfnamefont {Y.}~\bibnamefont {Lu}}, \bibinfo {author} {\bibfnamefont
  {H.}~\bibnamefont {Fang}}, \bibinfo {author} {\bibfnamefont {F.}~\bibnamefont
  {Sheng}}, \bibinfo {author} {\bibfnamefont {Y.}~\bibnamefont {Zhou}},
  \bibinfo {author} {\bibfnamefont {X.}~\bibnamefont {Yang}}, \bibinfo {author}
  {\bibfnamefont {Y.}~\bibnamefont {Li}}, \bibinfo {author} {\bibfnamefont
  {C.}~\bibnamefont {Feng}}, \ and\ \bibinfo {author} {\bibfnamefont {Z.-A.}\
  \bibnamefont {Xu}},\ }\href {\doibase 10.1103/PhysRevB.93.121112} {\bibfield
  {journal} {\bibinfo  {journal} {Phys. Rev. B}\ }\textbf {\bibinfo {volume}
  {93}},\ \bibinfo {pages} {121112} (\bibinfo {year}
  {2016}{\natexlab{b}})}\BibitemShut {NoStop}%
\bibitem [{\citenamefont {Shekhar}\ \emph {et~al.}(2015)\citenamefont
  {Shekhar}, \citenamefont {Nayak}, \citenamefont {Sun}, \citenamefont
  {Schmidt}, \citenamefont {Nicklas}, \citenamefont {Leermakers}, \citenamefont
  {Zeitler} \emph {et~al.}}]{XLMRNbpShekhar2015}%
  \BibitemOpen
  \bibfield  {author} {\bibinfo {author} {\bibfnamefont {C.}~\bibnamefont
  {Shekhar}}, \bibinfo {author} {\bibfnamefont {A.~K.}\ \bibnamefont {Nayak}},
  \bibinfo {author} {\bibfnamefont {Y.}~\bibnamefont {Sun}}, \bibinfo {author}
  {\bibfnamefont {M.}~\bibnamefont {Schmidt}}, \bibinfo {author} {\bibfnamefont
  {M.}~\bibnamefont {Nicklas}}, \bibinfo {author} {\bibfnamefont
  {I.}~\bibnamefont {Leermakers}}, \bibinfo {author} {\bibnamefont {Zeitler}},
  \emph {et~al.},\ }\href {http://dx.doi.org/10.1038/nphys3372} {\bibfield
  {journal} {\bibinfo  {journal} {Nat. Phys.}\ }\textbf {\bibinfo {volume}
  {11}},\ \bibinfo {pages} {645} (\bibinfo {year} {2015})}\BibitemShut
  {NoStop}%
\bibitem [{\citenamefont {Huang}\ \emph
  {et~al.}(2015{\natexlab{b}})\citenamefont {Huang}, \citenamefont {Zhao},
  \citenamefont {Long}, \citenamefont {Wang}, \citenamefont {Chen},
  \citenamefont {Yang}, \citenamefont {Liang} \emph
  {et~al.}}]{ChiralAnomalyTaAs_HuangPRX2015}%
  \BibitemOpen
  \bibfield  {author} {\bibinfo {author} {\bibfnamefont {X.}~\bibnamefont
  {Huang}}, \bibinfo {author} {\bibfnamefont {L.}~\bibnamefont {Zhao}},
  \bibinfo {author} {\bibfnamefont {Y.}~\bibnamefont {Long}}, \bibinfo {author}
  {\bibfnamefont {P.}~\bibnamefont {Wang}}, \bibinfo {author} {\bibfnamefont
  {D.}~\bibnamefont {Chen}}, \bibinfo {author} {\bibfnamefont {Z.}~\bibnamefont
  {Yang}}, \bibinfo {author} {\bibfnamefont {H.}~\bibnamefont {Liang}},  \emph
  {et~al.},\ }\href {\doibase 10.1103/PhysRevX.5.031023} {\bibfield  {journal}
  {\bibinfo  {journal} {Phys. Rev. X}\ }\textbf {\bibinfo {volume} {5}},\
  \bibinfo {pages} {031023} (\bibinfo {year} {2015}{\natexlab{b}})}\BibitemShut
  {NoStop}%
\bibitem [{\citenamefont {Bruno}\ \emph {et~al.}(2016)\citenamefont {Bruno},
  \citenamefont {Tamai}, \citenamefont {Wu}, \citenamefont {Cucchi},
  \citenamefont {Barreteau}, \citenamefont {de~la Torre}, \citenamefont
  {McKeown~Walker}, \citenamefont {Ricc\`o} \emph
  {et~al.}}]{Bruno_Baumberger_2016}%
  \BibitemOpen
  \bibfield  {author} {\bibinfo {author} {\bibfnamefont {F.~Y.}\ \bibnamefont
  {Bruno}}, \bibinfo {author} {\bibfnamefont {A.}~\bibnamefont {Tamai}},
  \bibinfo {author} {\bibfnamefont {Q.~S.}\ \bibnamefont {Wu}}, \bibinfo
  {author} {\bibfnamefont {I.}~\bibnamefont {Cucchi}}, \bibinfo {author}
  {\bibfnamefont {C.}~\bibnamefont {Barreteau}}, \bibinfo {author}
  {\bibfnamefont {A.}~\bibnamefont {de~la Torre}}, \bibinfo {author}
  {\bibfnamefont {S.}~\bibnamefont {McKeown~Walker}}, \bibinfo {author}
  {\bibfnamefont {S.}~\bibnamefont {Ricc\`o}},  \emph {et~al.},\ }\href
  {\doibase 10.1103/PhysRevB.94.121112} {\bibfield  {journal} {\bibinfo
  {journal} {Phys. Rev. B}\ }\textbf {\bibinfo {volume} {94}},\ \bibinfo
  {pages} {121112} (\bibinfo {year} {2016})}\BibitemShut {NoStop}%
\bibitem [{\citenamefont {McCormick}\ \emph {et~al.}(2017)\citenamefont
  {McCormick}, \citenamefont {Kimchi},\ and\ \citenamefont
  {Trivedi}}]{MinimalModelWeyl_Trivedi__RB2017}%
  \BibitemOpen
  \bibfield  {author} {\bibinfo {author} {\bibfnamefont {T.~M.}\ \bibnamefont
  {McCormick}}, \bibinfo {author} {\bibfnamefont {I.}~\bibnamefont {Kimchi}}, \
  and\ \bibinfo {author} {\bibfnamefont {N.}~\bibnamefont {Trivedi}},\ }\href
  {\doibase 10.1103/PhysRevB.95.075133} {\bibfield  {journal} {\bibinfo
  {journal} {Phys. Rev. B}\ }\textbf {\bibinfo {volume} {95}},\ \bibinfo
  {pages} {075133} (\bibinfo {year} {2017})}\BibitemShut {NoStop}%
\bibitem [{\citenamefont {Huang}\ \emph {et~al.}(2016)\citenamefont {Huang},
  \citenamefont {McCormick}, \citenamefont {Ochi}, \citenamefont {Zhao},
  \citenamefont {Suzuki}, \citenamefont {Arita}, \citenamefont {Wu} \emph
  {et~al.}}]{MoTe2Arpes_Kaminski_Nature2016}%
  \BibitemOpen
  \bibfield  {author} {\bibinfo {author} {\bibfnamefont {L.}~\bibnamefont
  {Huang}}, \bibinfo {author} {\bibfnamefont {T.~M.}\ \bibnamefont
  {McCormick}}, \bibinfo {author} {\bibfnamefont {M.}~\bibnamefont {Ochi}},
  \bibinfo {author} {\bibfnamefont {Z.}~\bibnamefont {Zhao}}, \bibinfo {author}
  {\bibfnamefont {M.-T.}\ \bibnamefont {Suzuki}}, \bibinfo {author}
  {\bibfnamefont {R.}~\bibnamefont {Arita}}, \bibinfo {author} {\bibnamefont
  {Wu}},  \emph {et~al.},\ }\href {http://dx.doi.org/10.1038/nmat4685}
  {\bibfield  {journal} {\bibinfo  {journal} {Nat. Mater.}\ }\textbf {\bibinfo
  {volume} {15}},\ \bibinfo {pages} {1155} (\bibinfo {year}
  {2016})}\BibitemShut {NoStop}%
\bibitem [{\citenamefont {Deng}\ \emph {et~al.}(2016)\citenamefont {Deng},
  \citenamefont {Wan}, \citenamefont {Deng}, \citenamefont {Zhang},
  \citenamefont {Ding}, \citenamefont {Wang}, \citenamefont {Yan},
  \citenamefont {Huang} \emph {et~al.}}]{MoTe2Arpes_Zhou_Nature2016}%
  \BibitemOpen
  \bibfield  {author} {\bibinfo {author} {\bibfnamefont {K.}~\bibnamefont
  {Deng}}, \bibinfo {author} {\bibfnamefont {G.}~\bibnamefont {Wan}}, \bibinfo
  {author} {\bibfnamefont {P.}~\bibnamefont {Deng}}, \bibinfo {author}
  {\bibfnamefont {K.}~\bibnamefont {Zhang}}, \bibinfo {author} {\bibfnamefont
  {S.}~\bibnamefont {Ding}}, \bibinfo {author} {\bibfnamefont {E.}~\bibnamefont
  {Wang}}, \bibinfo {author} {\bibfnamefont {M.}~\bibnamefont {Yan}}, \bibinfo
  {author} {\bibnamefont {Huang}},  \emph {et~al.},\ }\href
  {http://dx.doi.org/10.1038/nphys3871} {\bibfield  {journal} {\bibinfo
  {journal} {Nat. Phys.}\ }\textbf {\bibinfo {volume} {12}},\ \bibinfo {pages}
  {1105} (\bibinfo {year} {2016})}\BibitemShut {NoStop}%
\bibitem [{\citenamefont {Jiang}\ \emph {et~al.}(2017)\citenamefont {Jiang},
  \citenamefont {Liu}, \citenamefont {Sun}, \citenamefont {Yang}, \citenamefont
  {Rajamathi}, \citenamefont {Qi}, \citenamefont {Yang}, \citenamefont {Chen},
  \citenamefont {Peng} \emph {et~al.}}]{MoTe2Arpes_jiang_Nature2017}%
  \BibitemOpen
  \bibfield  {author} {\bibinfo {author} {\bibfnamefont {J.}~\bibnamefont
  {Jiang}}, \bibinfo {author} {\bibfnamefont {Z.}~\bibnamefont {Liu}}, \bibinfo
  {author} {\bibfnamefont {Y.}~\bibnamefont {Sun}}, \bibinfo {author}
  {\bibfnamefont {H.}~\bibnamefont {Yang}}, \bibinfo {author} {\bibfnamefont
  {C.}~\bibnamefont {Rajamathi}}, \bibinfo {author} {\bibfnamefont
  {Y.}~\bibnamefont {Qi}}, \bibinfo {author} {\bibfnamefont {L.}~\bibnamefont
  {Yang}}, \bibinfo {author} {\bibfnamefont {C.}~\bibnamefont {Chen}}, \bibinfo
  {author} {\bibfnamefont {H.}~\bibnamefont {Peng}},  \emph {et~al.},\ }\href
  {http://dx.doi.org/10.1038/ncomms13973} {\bibfield  {journal} {\bibinfo
  {journal} {Nat. Commun.}\ }\textbf {\bibinfo {volume} {8}},\ \bibinfo {pages}
  {13973} (\bibinfo {year} {2017})}\BibitemShut {NoStop}%
\bibitem [{\citenamefont {Tamai}\ \emph {et~al.}(2016)\citenamefont {Tamai},
  \citenamefont {Wu}, \citenamefont {Cucchi}, \citenamefont {Bruno},
  \citenamefont {Ricc\`o}, \citenamefont {Kim}, \citenamefont {Hoesch},
  \citenamefont {Barreteau} \emph {et~al.}}]{Tamai_Baumberger_2016}%
  \BibitemOpen
  \bibfield  {author} {\bibinfo {author} {\bibfnamefont {A.}~\bibnamefont
  {Tamai}}, \bibinfo {author} {\bibfnamefont {Q.~S.}\ \bibnamefont {Wu}},
  \bibinfo {author} {\bibfnamefont {I.}~\bibnamefont {Cucchi}}, \bibinfo
  {author} {\bibfnamefont {F.~Y.}\ \bibnamefont {Bruno}}, \bibinfo {author}
  {\bibfnamefont {S.}~\bibnamefont {Ricc\`o}}, \bibinfo {author} {\bibfnamefont
  {T.~K.}\ \bibnamefont {Kim}}, \bibinfo {author} {\bibfnamefont
  {M.}~\bibnamefont {Hoesch}}, \bibinfo {author} {\bibfnamefont
  {C.}~\bibnamefont {Barreteau}},  \emph {et~al.},\ }\href {\doibase
  10.1103/PhysRevX.6.031021} {\bibfield  {journal} {\bibinfo  {journal} {Phys.
  Rev. X}\ }\textbf {\bibinfo {volume} {6}},\ \bibinfo {pages} {031021}
  (\bibinfo {year} {2016})}\BibitemShut {NoStop}%
\bibitem [{\citenamefont {Rhodes}\ \emph {et~al.}(2017)\citenamefont {Rhodes},
  \citenamefont {Sch\"onemann}, \citenamefont {Aryal}, \citenamefont {Zhou},
  \citenamefont {Zhang} \emph {et~al.}}]{MoTe2Aryal2017}%
  \BibitemOpen
  \bibfield  {author} {\bibinfo {author} {\bibfnamefont {D.}~\bibnamefont
  {Rhodes}}, \bibinfo {author} {\bibfnamefont {R.}~\bibnamefont
  {Sch\"onemann}}, \bibinfo {author} {\bibfnamefont {N.}~\bibnamefont {Aryal}},
  \bibinfo {author} {\bibfnamefont {Q.}~\bibnamefont {Zhou}}, \bibinfo {author}
  {\bibfnamefont {Q.~R.}\ \bibnamefont {Zhang}},  \emph {et~al.},\ }\href
  {\doibase 10.1103/PhysRevB.96.165134} {\bibfield  {journal} {\bibinfo
  {journal} {Phys. Rev. B}\ }\textbf {\bibinfo {volume} {96}},\ \bibinfo
  {pages} {165134} (\bibinfo {year} {2017})}\BibitemShut {NoStop}%
\bibitem [{\citenamefont {Sun}\ \emph {et~al.}(2015)\citenamefont {Sun},
  \citenamefont {Wu}, \citenamefont {Ali}, \citenamefont {Felser},\ and\
  \citenamefont {Yan}}]{MoTe2YanPRB2015}%
  \BibitemOpen
  \bibfield  {author} {\bibinfo {author} {\bibfnamefont {Y.}~\bibnamefont
  {Sun}}, \bibinfo {author} {\bibfnamefont {S.-C.}\ \bibnamefont {Wu}},
  \bibinfo {author} {\bibfnamefont {M.~N.}\ \bibnamefont {Ali}}, \bibinfo
  {author} {\bibfnamefont {C.}~\bibnamefont {Felser}}, \ and\ \bibinfo {author}
  {\bibfnamefont {B.}~\bibnamefont {Yan}},\ }\href {\doibase
  10.1103/PhysRevB.92.161107} {\bibfield  {journal} {\bibinfo  {journal} {Phys.
  Rev. B}\ }\textbf {\bibinfo {volume} {92}},\ \bibinfo {pages} {161107}
  (\bibinfo {year} {2015})}\BibitemShut {NoStop}%
\bibitem [{\citenamefont {Crepaldi}\ \emph {et~al.}(2017)\citenamefont
  {Crepaldi}, \citenamefont {Aut\`es}, \citenamefont {Sterzi}, \citenamefont
  {Manzoni}, \citenamefont {Zacchigna}, \citenamefont {Cilento} \emph
  {et~al.}}]{TrivialMoTe2_Crepaldi_Grioni_PRB2017}%
  \BibitemOpen
  \bibfield  {author} {\bibinfo {author} {\bibfnamefont {A.}~\bibnamefont
  {Crepaldi}}, \bibinfo {author} {\bibfnamefont {G.}~\bibnamefont {Aut\`es}},
  \bibinfo {author} {\bibfnamefont {A.}~\bibnamefont {Sterzi}}, \bibinfo
  {author} {\bibfnamefont {G.}~\bibnamefont {Manzoni}}, \bibinfo {author}
  {\bibfnamefont {M.}~\bibnamefont {Zacchigna}}, \bibinfo {author}
  {\bibfnamefont {F.}~\bibnamefont {Cilento}},  \emph {et~al.},\ }\href
  {\doibase 10.1103/PhysRevB.95.041408} {\bibfield  {journal} {\bibinfo
  {journal} {Phys. Rev. B}\ }\textbf {\bibinfo {volume} {95}},\ \bibinfo
  {pages} {041408} (\bibinfo {year} {2017})}\BibitemShut {NoStop}%
\bibitem [{Sup()}]{Supplementary}%
  \BibitemOpen
  \href@noop {} {}\bibinfo {note} {See Supplemental Material}\BibitemShut
  {NoStop}%
\bibitem [{\citenamefont {Qi}\ \emph {et~al.}(2016)\citenamefont {Qi},
  \citenamefont {Naumov}, \citenamefont {Ali}, \citenamefont {Rajamathi},
  \citenamefont {Schnelle} \emph {et~al.}}]{MoTe2HSE_Qi_Nature2016}%
  \BibitemOpen
  \bibfield  {author} {\bibinfo {author} {\bibfnamefont {Y.}~\bibnamefont
  {Qi}}, \bibinfo {author} {\bibfnamefont {P.~G.}\ \bibnamefont {Naumov}},
  \bibinfo {author} {\bibfnamefont {M.~N.}\ \bibnamefont {Ali}}, \bibinfo
  {author} {\bibfnamefont {C.~R.}\ \bibnamefont {Rajamathi}}, \bibinfo {author}
  {\bibfnamefont {W.}~\bibnamefont {Schnelle}},  \emph {et~al.},\ }\href
  {http://dx.doi.org/10.1038/ncomms11038} {\bibfield  {journal} {\bibinfo
  {journal} {Nat. Commun.}\ }\textbf {\bibinfo {volume} {7}},\ \bibinfo {pages}
  {11038} (\bibinfo {year} {2016})}\BibitemShut {NoStop}%
\bibitem [{\citenamefont {\L{}usakowski}\ \emph {et~al.}(2011)\citenamefont
  {\L{}usakowski}, \citenamefont {Bogus\l{}awski},\ and\ \citenamefont
  {Radzy\ifmmode~\acute{n}\else \'{n}\fi{}ski}}]{RoleSOI_Radzynski2011}%
  \BibitemOpen
  \bibfield  {author} {\bibinfo {author} {\bibfnamefont {A.}~\bibnamefont
  {\L{}usakowski}}, \bibinfo {author} {\bibfnamefont {P.}~\bibnamefont
  {Bogus\l{}awski}}, \ and\ \bibinfo {author} {\bibfnamefont {T.}~\bibnamefont
  {Radzy\ifmmode~\acute{n}\else \'{n}\fi{}ski}},\ }\href {\doibase
  10.1103/PhysRevB.83.115206} {\bibfield  {journal} {\bibinfo  {journal} {Phys.
  Rev. B}\ }\textbf {\bibinfo {volume} {83}},\ \bibinfo {pages} {115206}
  (\bibinfo {year} {2011})}\BibitemShut {NoStop}%
\bibitem [{\citenamefont {Sch\"onemann}\ \emph {et~al.}(2017)\citenamefont
  {Sch\"onemann}, \citenamefont {Aryal}, \citenamefont {Zhou}, \citenamefont
  {Chiu}, \citenamefont {Chen}, \citenamefont {Martin} \emph
  {et~al.}}]{WP2_Schonemann_Aryal_PRB_2017}%
  \BibitemOpen
  \bibfield  {author} {\bibinfo {author} {\bibfnamefont {R.}~\bibnamefont
  {Sch\"onemann}}, \bibinfo {author} {\bibfnamefont {N.}~\bibnamefont {Aryal}},
  \bibinfo {author} {\bibfnamefont {Q.}~\bibnamefont {Zhou}}, \bibinfo {author}
  {\bibfnamefont {Y.-C.}\ \bibnamefont {Chiu}}, \bibinfo {author}
  {\bibfnamefont {K.-W.}\ \bibnamefont {Chen}}, \bibinfo {author}
  {\bibfnamefont {T.~J.}\ \bibnamefont {Martin}},  \emph {et~al.},\ }\href
  {\doibase 10.1103/PhysRevB.96.121108} {\bibfield  {journal} {\bibinfo
  {journal} {Phys. Rev. B}\ }\textbf {\bibinfo {volume} {96}},\ \bibinfo
  {pages} {121108} (\bibinfo {year} {2017})}\BibitemShut {NoStop}%
\bibitem [{\citenamefont {Chen}\ \emph {et~al.}(2018)\citenamefont {Chen},
  \citenamefont {Lian}, \citenamefont {Lai}, \citenamefont {Aryal},
  \citenamefont {Chiu} \emph {et~al.}}]{MAl3_KWChen_PRL2018}%
  \BibitemOpen
  \bibfield  {author} {\bibinfo {author} {\bibfnamefont {K.-W.}\ \bibnamefont
  {Chen}}, \bibinfo {author} {\bibfnamefont {X.}~\bibnamefont {Lian}}, \bibinfo
  {author} {\bibfnamefont {Y.}~\bibnamefont {Lai}}, \bibinfo {author}
  {\bibfnamefont {N.}~\bibnamefont {Aryal}}, \bibinfo {author} {\bibfnamefont
  {Y.-C.}\ \bibnamefont {Chiu}},  \emph {et~al.},\ }\href {\doibase
  10.1103/PhysRevLett.120.206401} {\bibfield  {journal} {\bibinfo  {journal}
  {Phys. Rev. Lett.}\ }\textbf {\bibinfo {volume} {120}},\ \bibinfo {pages}
  {206401} (\bibinfo {year} {2018})}\BibitemShut {NoStop}%
\bibitem [{\citenamefont {Di~Sante}\ \emph {et~al.}(2017)\citenamefont
  {Di~Sante}, \citenamefont {Das}, \citenamefont {Bigi}, \citenamefont
  {Erg\"onenc}, \citenamefont {G\"urtler}, \citenamefont {Krieger} \emph
  {et~al.}}]{CorrelationsWTe2_Di_Panaccione_PRL_2017}%
  \BibitemOpen
  \bibfield  {author} {\bibinfo {author} {\bibfnamefont {D.}~\bibnamefont
  {Di~Sante}}, \bibinfo {author} {\bibfnamefont {P.~K.}\ \bibnamefont {Das}},
  \bibinfo {author} {\bibfnamefont {C.}~\bibnamefont {Bigi}}, \bibinfo {author}
  {\bibfnamefont {Z.}~\bibnamefont {Erg\"onenc}}, \bibinfo {author}
  {\bibfnamefont {N.}~\bibnamefont {G\"urtler}}, \bibinfo {author}
  {\bibnamefont {Krieger}},  \emph {et~al.},\ }\href {\doibase
  10.1103/PhysRevLett.119.026403} {\bibfield  {journal} {\bibinfo  {journal}
  {Phys. Rev. Lett.}\ }\textbf {\bibinfo {volume} {119}},\ \bibinfo {pages}
  {026403} (\bibinfo {year} {2017})}\BibitemShut {NoStop}%
\end{thebibliography}
\end{document}